\documentclass[journal]{IEEEtran}
\usepackage{amsmath,amssymb}
\usepackage{graphicx}
\usepackage{booktabs}
\usepackage{multirow}
\usepackage{array}

\usepackage[
    caption=false,
    font=footnotesize,
    labelfont=sf,
    textfont=sf
]{subfig}

\usepackage{xcolor}

\usepackage[nocompress]{cite}
\usepackage[hidelinks]{hyperref}

\ifCLASSINFOpdf
\else
\fi
\begin{document}
%
\title{CGSpec: Consistency-Guided Speculative Decoding with Overlap-Aware Verification Scheduling for Edge–Cloud Inference}
%
%
%


\author{
Yuanbing Ouyang,~\IEEEmembership{Member,~IEEE},
Xiaoli Liu,~\IEEEmembership{Member,~IEEE},
Shirong Wei,
Hao Wang,~\IEEEmembership{Senior Member,~IEEE}, 
and Xiang Su,~\IEEEmembership{Member,~IEEE}, \\

\thanks{Yuanbing Ouyang, Shirong Wei and Hao Wang are with the School of Cyber Engineering, Xidian University, xi'an, china. E-mail:yuanbing.oy@stu.xidian.edu.cn, srwei@stu.xidian.edu.cn and haow@ieee.org}
\thanks{X Liu is with School of Technology and Innovations, University of Vaasa and Department of Computer Science, University of Helsinki, Finland. E-mail: xiaoli.liu@uwasa.fi}
\thanks{X. Su is with the with Department of Computer Science and Department of Agricultural Sciences, University of Helsinki, 00014 Helsinki, Finland. E-mail: xiang.su@helsinki.fi}

\thanks{Manuscript received November 15, 2026; revised N, 2027.}}

%
%

\markboth{}{}
%



\maketitle

\title{
CGSpec: Consistency-Guided Speculative Decoding
with Overlap-Aware Verification Scheduling
for edge-cloud continuum
}

\begin{abstract}
Collaborative speculative decoding accelerates large language model inference by enabling edge clients to draft tokens locally and a shared cloud model to verify them. However, draft–target disagreement wastes communication and verification resources, while heterogeneous client return times complicate shared-verifier scheduling. We propose CGSpec, a consistency guided framework for multi client edge–cloud speculative decoding. CGSpec uses draft–target distribution divergence to adapt the speculative budget and introduces a nomination and competition mechanism to construct candidate trees through global growth control and local token selection. An overlap-aware scheduler further estimates remaining verification work and client return times to prioritize requests and improve verifier utilization. We evaluate CGSpec using four model pairs and three datasets under heterogeneous network conditions. Across 24 configurations with 8 and 16 clients, CGSpec achieves the highest mean system and client throughput among the evaluated methods. Compared with current SOTA methods, it improves system throughput by 12.2\% and 10.2\%, and mean client throughput by 14.0\% and 33.0\%, respectively.
\end{abstract}


\IEEEpeerreviewmaketitle

\section{Introduction}

Large language models (LLMs) are increasingly deployed in networked sensing systems, where interactive services require low latency, high quality, and efficient collaboration between edge devices and cloud infrastructure~\cite{ubicomp513}.
Emerging latency sensitive applications impose strong requirements on LLM inference systems, which must balance response time, device constraints, and edge-cloud communication overhead.
However, autoregressive decoding remains a major bottleneck because tokens must be generated sequentially and each step requires a full forward pass over the current prefix. This bottleneck is especially severe on resource constrained edge devices, which have limited computation, memory bandwidth, and energy budgets.

Speculative decoding (SD) accelerates autoregressive generation by allowing a lightweight draft model to propose multiple candidate tokens, which are then verified in parallel by a stronger target model while preserving the target distribution~\cite{leviathan2023specdec, chen2023specsampling}. Despite its effectiveness in accelerating LLM inference, deploying SD on edge devices remains challenging because the target model typically requires substantial computation and memory resources. To overcome this limitation, collaborative SD leverages the complementary capability of edge devices and cloud infrastructure, where edge devices perform lightweight local drafting, and the cloud verifies candidates with a larger target model~\cite{park2026specedge}. Recent systems have shown that this design can improve throughput by combining low latency edge-side drafting with high capacity cloud verification~\cite{li2025sled, park2026specedge, 10812936, yu2025dsd}.

Despite this progress, achieving consistent and efficient collaborative SD in heterogeneous network environments remains challenge.
The quality of the drafter and the speed of the verifier can no longer be considered in isolation. The candidate drafts generated on the edge are no longer free internal candidates; they consume uplink bandwidth, verifier computation, and part of the shared verification opportunity when submitted to the cloud. Meanwhile, verification requests from different clients arrive irregularly due to heterogeneous network delays, context length, cache footprint, and verification behavior.
These factors introduce two key challenges.

First, determining an appropriate speculative budget for
each request under varying draft–target agreement. Existing adaptive methods usually regulate speculation using draft uncertainty alone, for example by adjusting lookahead length, pruning low confidence candidates, or dynamically reshaping trees according to draft confidence~\cite{mamou2024disco, li2024eagle, xiong2025dyspec, chen2025sequoia, guan2026yggdrasil}. However, this design mixes two factors that should be considered separately. The amount of speculation that is worth verifying is determined primarily by the agreement between the draft and target models, and this agreement can vary substantially across tasks and prompts. By contrast, the way a given budget should be organized inside the draft tree is determined by the local uncertainty of the draft model. When these two roles are conflated, the system may expand too aggressively under poor draft-target alignment and waste communication and verification budget, or behave too conservatively under high consistency and fail to exploit cloud parallelism.


\textbf{Second, coordinating shared verification with heterogeneous client return paths} Multiple clients share a target model for candidate verification. Each client alternates between local drafting, communication, and cloud verification. Serving a request from a ready client determines not only the current verification order but also when that client will return with its next batch of candidates. Consequently, prioritizing a client with a longer drafting and communication cycle may delay its next request, potentially leaving the cloud verifier idle after other clients have completed their available work. Although existing systems exploit pipeline overlap, such as SpecEdge's interleaving of user requests~\cite{park2026specedge}, selecting the next client remains challenging when clients have heterogeneous drafting and communication costs. Our scheduling question is how to leverage observed return costs and remaining verification workloads to select among ready clients and prevent idle periods.

These two challenges point to a holistic systems question: \textit{in multi-client collaborative SD between edge and cloud, how should the system allocate limited speculative budget within each request and limited verification opportunities across requests so as to maximize end-to-end throughput?} This requires joint control over communication efficiency, verification amortization, and queueing under shared verifier contention.

\textbf{CGSpec}, a consistency-guided tree construction with overlap-aware verification scheduling framework for multi-client collaborative SD between edge and cloud. CGSpec is built around a three-stage control design. First, it introduces a consistency-guided state sensing mechanism that estimates draft-target agreement from verification feedback using Jensen Shannon (JS) divergence, and uses this signal to regulate how much speculative budget should be allocated to the current round. Second, under the consistency conditioned budget, it constructs speculative trees through a nomination and competition process that separates global growth control from local token nomination, thereby improving accepted progress per verification round without over expanding the draft tree. Third, under concurrent deployment, it allocates cloud verification opportunities through a cost aware scheduler that jointly accounts for waiting urgency, expected effective gain and serving cost, while using anti-starvation protection only as a lightweight safeguard.

The core contributions of this work are summarized as
follows:
\begin{itemize}

\item
We formulate a coupled optimization problem for multi client edge–cloud speculative decoding, characterizing the interactions among local drafting, network communication, and shared-server verification under heterogeneous client conditions. 

\item 
We design a consistency-guided tree construction mechanism that uses draft–target distribution divergence to adapt speculative budgets, and introduces a nomination and competition strategy to select globally valuable candidates under limited resources. 

\item 
We introduce an overlap-aware verification scheduling mechanism that estimates remaining workloads and client return times to prioritize verification requests, improving shared-server utilization while protecting client waiting time.

\item
We conduct extensive experiments covering four model pairs, three datasets, heterogeneous 4G/5G networks, and up to 30 clients. Comprehensive ablation studies and latency analyses further validate the effectiveness of each proposed mechanism.

\end{itemize}
\section{Related Work}
\label{sec:relatedwork}

\subsection{Speculative Decoding}

SD has expanded into a broad area covering drafter design, online control, tree-structured verification, and distributed serving since its original draft and verify formulation \cite{xia2024survey}. Existing methods mainly differ in how candidates are produced, how they are organized for verification, and which signals regulate speculation. Most prior work on SD focuses on improving the quality of draft candidates or the control of speculative budget. On the drafting side, Staged SD organizes proposals hierarchically across multiple drafting stages \cite{spector2023staged} while EAGLE and EAGLE-3 improve the drafting objective through feature level or direct token prediction \cite{li2024eagle,li2026eagle}. AdaSpec strengthens draft--target alignment through selective knowledge distillation \cite{hu2026adaspec}. On the budget control side, DISCO dynamically adjusts lookahead length instead of using a static speculative budget \cite{mamou2024disco}. These studies establish SD as a mature exact decoding paradigm, but they are primarily designed for centralized inference, where verifier acceleration dominates the objective and deployment costs are secondary. In collaborative edge-cloud inference, draft-target consistency can vary across tasks, while uplink delay and verifier occupancy become first-order factors in end-to-end performance \cite{li2025sled,park2026specedge,tran2025goodspeed,yu2025dsd}.

\subsection{Tree-Based SD}

Early SD formulations mostly follow a chain based verification pattern: the drafter proposes a linear continuation and the verifier checks the sequence in one step \cite{leviathan2023specdec,chen2023specsampling}. Chain verification is sensitive to early mismatch because one incorrect token can invalidate much of the remaining draft \cite{leviathan2023specdec,spector2023staged}. This limitation motivated tree based SD, where richer candidate structures expose more parallelism during verification. SpecInfer verifies multiple paths in a token tree \cite{miao2024specinfer}, while Medusa generates tree structured candidates with multiple decoding heads on the backbone model \cite{cai2024medusa}. Recent methods have shifted from tree based verification to optimized tree construction itself. Sequoia uses dynamic programming and hardware-aware optimization to select tree size and depth \cite{chen2025sequoia}. EAGLE-2 builds context-sensitive draft trees using confidence in the draft model \cite{li2024eagle2}. DySpec reconstructs tree topology online under changing contexts \cite{xiong2025dyspec} and Yggdrasil co-designs tree growth with runtime execution for latency optimal decoding \cite{guan2026yggdrasil}. These methods show that verification efficiency depends not only on the number of proposed tokens, but also on how candidate tree is structured before verification.

However, prior tree-based methods remain mainly draft centric. They usually use draft-side confidence to prune low confidence candidates or decide whether expansion should continue \cite{li2024eagle2,xiong2025dyspec}. Such signals are useful for local exploration, but they do not directly answer whether the resulting token set is globally worthwhile under an upload and verification budget. In collaborative SD, drafted tokens are not free internal candidates: each extra token consumes uplink bandwidth, cloud verification computation, and part of the shared verification opportunity. Tree construction should therefore be treated not only as local confidence-guided expansion, but also as budgeted selection over globally valuable candidates.

\subsection{Collaborative SD between edge and cloud}

Recent research extends SD from single device acceleration to heterogeneous distributed environments. SLED lets heterogeneous edge devices draft locally while a shared server verifies requests in batches \cite{li2025sled}. SpecEdge splits work between edge and server GPUs and combines proactive edge drafting with pipeline-aware scheduling \cite{park2026specedge}. DSSD reduces uplink communication by splitting verification between device and edge \cite{ning2025dssd}, while Jupiter studies collaborative generative inference over decentralized edge resources \cite{ye2025jupiter}. These systems position the edge as an active participant in SD rather than a thin client. A growing body of work focuses on optimizing distributed SD execution such as orchestration, overlap, and communication efficiency. DSD coordinates draft-target execution across multiple devices through simulation driven control \cite{yu2025dsd}. FlowSpec and AMUSD improve overlap through pipelined or asynchronous execution \cite{liu2025flowspec,mcdanel2024amusd}. GoodSpeed and WISP study batching and budget design under objectives such as goodput, fairness, or SLOs \cite{tran2025goodspeed,li2026wisp}, and QSV plus Conformal Sparsification treat communication cost as a first-class factor in edge-cloud SD \cite{zhang2026qsv,bhattacharjee2025conformal}.


Existing edge-cloud SD systems improve resource utilization through pipelining, asynchronous execution, and batched verification. When multiple clients share a verifier, differences in drafting capability and network conditions lead to different return times between consecutive verification rounds. The verification order therefore affects both current waiting times and the arrival of subsequent verification requests. This creates an additional scheduling challenge: verification decisions must account for both the immediate service cost and the overlap between ongoing verification tasks and subsequent client return cycles.

\section{Problem Formulation}
\label{sec:formulation}

\subsection{System Model}

Consider a collaborative SD system with
$N$ clients and a cloud verification server.
Each client runs a lightweight draft model, while the cloud server hosts the target model for candidate verification.
Clients may be deployed on edge GPUs or personal GPU servers,
with heterogeneous local computing capacities and network conditions.
Each client maintains its generation prefix, constructs a candidate token tree locally,
and submits it to the cloud for verification.%

For client $i$ at decoding round $r$, let $x_{i,r}$ denote
the current generation prefix.
The next token distributions of the draft and target models are
$p_d(\cdot\mid x_{i,r})$ and $p_t(\cdot\mid x_{i,r})$, respectively.
The candidate tree is controlled by four structural limits,
$(K_{i,r},L_{i,r},W_{i,r},B_{i,r})$:
the maximum number of active paths, maximum tree depth,
maximum local branching width per parent,
and total node budget.
These parameters determine the organization of speculative candidates
and the work required for local drafting, uploading,
and cloud verification.%

Let $\mathcal A(t)$ denote the set of active clients at time $t$,
and let $\mathcal Q(t)$ denote the set of requests that have arrived
at the server and are ready for dispatch.
An active client may be drafting, communicating, waiting,
or receiving verification.
Consequently, an active client is not necessarily ready for service:
the scheduler can select only from $\mathcal Q(t)$.%

We first consider a server that executes one non preemptive
verification request at a time, matching the single request
verification setting of our scheduling design.
Local drafting can proceed concurrently across clients,
while cloud verification capacity is shared.

\subsection{Client Drafting and the Decoding Cycle}

Let $F_{i,r}$ denote the time required by client $i$
to construct its candidate tree in round $r$.
Drafting includes draft token generation, candidate selection,
and tree organization.
We express its cost as%

{
\begin{equation}
F_{i,r}
=
\sum_{\ell=1}^{L_{i,r}}
f_i(n_{i,r,\ell},c_{i,r,\ell})
+
h_{i,r},
\label{eq:draft-cost}
\end{equation}

where $n_{i,r,\ell}$ is the number of active prefixes
actually evaluated at expansion step $\ell$,
$c_{i,r,\ell}$ is their context length,
and $f_i$ is the corresponding draft token generation time
on client $i$.
The term $h_{i,r}$ includes candidate selection,
tree maintenance, and packing overhead.
Expansion steps that are not executed contribute zero cost.%

This representation allows multiple prefixes to be expanded
in a local batch, so drafting time need not scale linearly
with the final number of tree nodes.
The scheduler uses observed execution times
and does not require a fitted analytical form for $f_i$.%

Each client follows a sequential
draft, upload, verify then feedback cycle.
After receiving verification feedback,
the client updates its generation prefix
and starts the next drafting round.
Drafting and communication at one client can proceed concurrently
with the server's verification of other clients.%

For a request with $m_i$ verification rounds,
let $U_{i,r}$, $Q_{i,r}$, $S_{i,r}$, and $V_{i,r}$
denote its upload time, server queueing time,
full verification duration, and feedback time
in round $r$, respectively.
Its completion latency is%

{
\begin{equation}
\begin{aligned}
T_i
&=
T_i^{\mathrm{init}}
+
\sum_{r=1}^{m_i}
\bigl(
F_{i,r}
+
U_{i,r}
+
Q_{i,r}\\
&\hspace{6em}
+
S_{i,r}
+
V_{i,r}
\bigr).
\end{aligned}
\label{eq:request-latency}
\end{equation}
}

Here, $T_i^{\mathrm{init}}$ includes request initialization
and prefill outside the decode rounds.
The term $S_{i,r}$ measures the full serialized server slot
occupied by the request, including server processing
beyond the target device forward pass.
The stages use disjoint timing boundaries
to avoid double counting.%

Between consecutive rounds of an unfinished request,
the client receives feedback, constructs the next candidate tree,
and uploads it.
We define the return time from verification completion
to the next tree arrival as%

{
\begin{equation}
R_{i,r}
=
V_{i,r}
+
F_{i,r+1}
+
U_{i,r+1}.
\label{eq:return-cost}
\end{equation}
}

While client $i$ traverses this return path,
the server can verify requests from other clients.
Local drafting and communication therefore postpone
the next time client $i$ becomes ready,
but need not cause an equal amount of server idle time.
Their effect depends on whether other verification work
is available during that interval.%

Expanding a candidate tree can increase drafting,
upload, and verification costs,
while potentially reducing the number of rounds
through greater output progress per verification.
Dispatch order determines when each client starts
its next return cycle and how much of that cycle
can overlap verification work from other clients.
Tree construction and verification scheduling
thus jointly determine end-to-end performance.%

\subsection{Performance Objectives and Online Information}

For a finite workload measured from a common start time $t_0$
until completion at $t_1$, system throughput is%
}
{
\begin{equation}
\Gamma_{\mathrm{sys}}
=
\frac{\sum_{i=1}^{N}Y_i}{t_1-t_0},
\label{eq:system-throughput}
\end{equation}

where $Y_i$ is the number of tokens actually emitted
by client $i$ during the measurement phase.
The mean client generation rate is%
}
{
\begin{equation}
\Gamma_{\mathrm{client}}
=
\frac{1}{N}
\sum_{i=1}^{N}
\frac{Y_i}{T_i^{\mathrm{session}}},
\label{eq:client-throughput}
\end{equation}

where $T_i^{\mathrm{session}}$ is the time taken by client $i$
to complete its corresponding measured workload.%

These metrics capture different aspects of performance.
System throughput measures the completion efficiency
of the entire workload,
whereas the mean client generation rate measures
the output rate experienced by individual clients.
Improving the mean client rate does not necessarily
shorten the overall workload completion time.
We therefore evaluate both metrics,
together with request latency and waiting time distributions.%

Whenever a verification slot becomes available,
the scheduler selects the next request from $\mathcal Q(t)$,
subject to the execution and memory constraints
of the serving backend.
Its decision uses only information available at that time,
including request  processing,
actual emitted token counts,
completed request lengths,
observed service and return times,
and statistics of the current candidate trees.%

Under these constraints, we study two coupled online decisions.
At each client, draft-target consistency guides
the allocation of the tree budget,
controlling the execution cost and useful output of each round.
At the server, verification opportunities are allocated
according to ready request states,
estimated remaining verification work,
and client return times.
The objective is to improve the utilization
of shared verification capacity and client generation progress
under unknown output lengths
and heterogeneous computing and network conditions.%

\begin{figure*}[t]
    \centering
    \includegraphics[width=\textwidth]{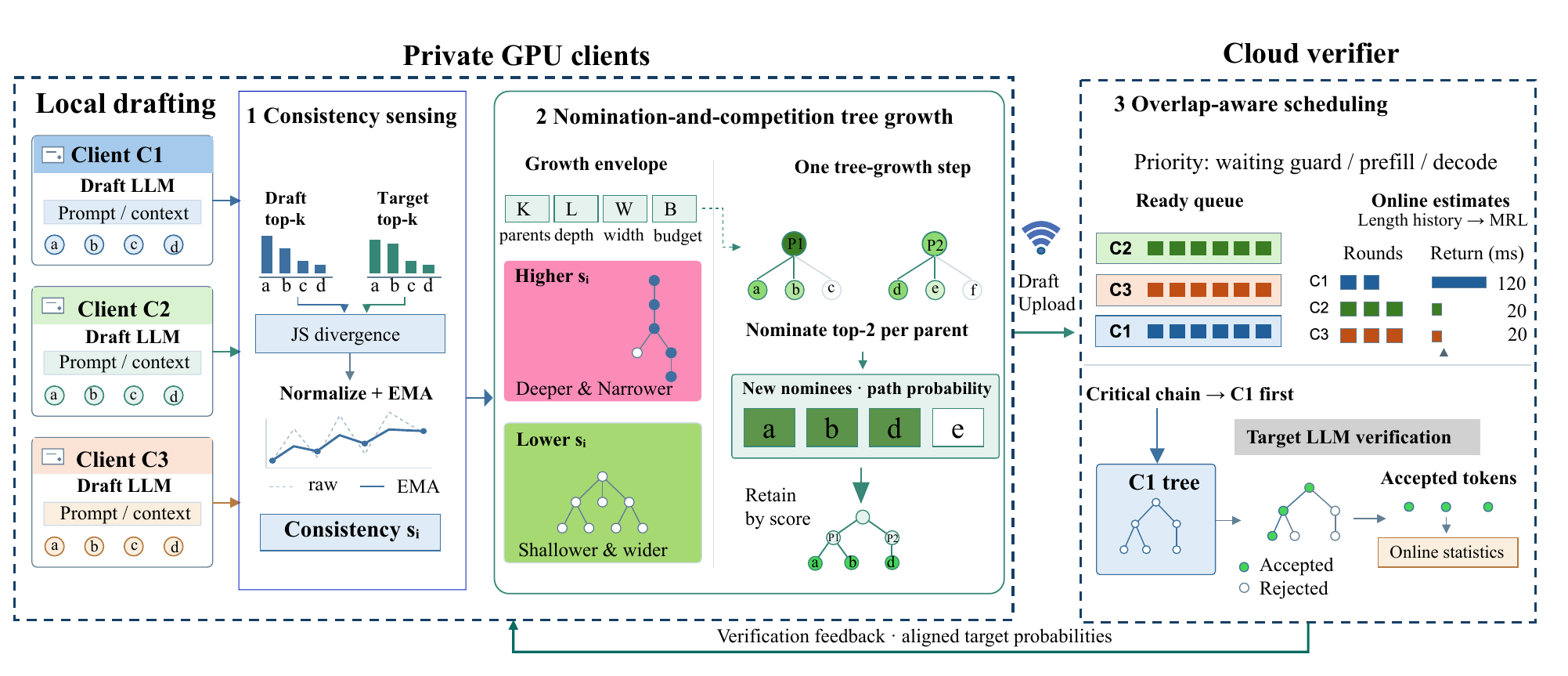}
    \caption{
    Overview of CGSpec.
    Private GPU clients use verification feedback to estimate
    draft-target consistency and control the growth envelope
    of a nomination and competition draft tree.
    The cloud verifier maintains a ready queue and online estimates
    of remaining verification work and client return times.
    An overlap-aware scheduler selects the next request,
    and verification feedback updates both the local generation
    state and subsequent control decisions.%
    }
    \label{fig:system}
\end{figure*}

\section{Methodology}
\label{sec:methodology}

\subsection{CGSpec Framework}

Figure~\ref{fig:system} presents the execution and control flow
of CGSpec.
Each client constructs a candidate tree with its local draft model
and uploads it to the shared cloud verifier.
After verification, the client receives the accepted output
and target model probability feedback,
updates its prefix, and begins the next drafting round.
Different clients can draft and communicate
while the server verifies another client's request.%

CGSpec coordinates this process through three components.
Consistency-guided state sensing compares aligned draft
and target distributions using normalized JS
divergence and smooths the resulting signal over time
(Section~\ref{sec:CGSS}).
nomination and competition tree growth uses this state
to set a growth envelope and select candidates
within the node budget (Section~\ref{sec:NCTree}). 
Overlap-aware scheduling combines waiting protection,
prefill priority, and decode selection based on remaining
verification work and client return times
(Section~\ref{Sec:CAS}).%

These components act on the same decoding cycle.
Tree construction determines local drafting cost,
upload volume, verification cost,
and output progress per round.
Scheduling determines when each client starts its next
feedback draft upload cycle and how that cycle overlaps
verification work from other clients.
Figure~\ref{fig:scheduling-overlap} illustrates this interaction
through a concrete scheduling example.%

\subsection{Consistency-Guided State Sensing}
\label{sec:CGSS}

A central question in adaptive SD concerns the allocation
of the draft budget to the current round.
This issue is especially important in SD between edge and cloud,
where drafted tokens are not free:
every additional token submitted by the edge incurs extra
communication and cloud-side verification overhead.
Therefore, instead of using draft model's uncertainty only
as a local token score, we require a global control signal
that can determine whether the overall speculative budget
of the current round is worthwhile for verification.

To this end, we compare the draft and target next-token
distributions on an aligned top support $k$ 
and quantify their discrepancy using JS divergence.
Let $\tilde{p}_d(\cdot\mid x_t)$ and
$\tilde{p}_t(\cdot\mid x_t)$ denote the normalized top-$k$
distributions of the draft and target models
under the current prefix $x_t$.
Their mismatch is defined as
\begin{equation}
\begin{aligned}
J_t
&=JS(\tilde p_d\|\tilde p_t)\\
&=\tfrac12 KL(\tilde p_d\|M_t)
 +\tfrac12 KL(\tilde p_t\|M_t),\\
M_t
&=\tfrac12(\tilde p_d+\tilde p_t).
\end{aligned}
\label{eq:js-divergence}
\end{equation}

Here, JS divergence is used to characterize the current draft–target agreement rather than to determine token acceptance.
Compared with draft-only uncertainty,
this signal is more directly coupled to effective verification gain;
compared with discrete acceptance statistics,
it preserves the local probability structure
and can distinguish mild disagreement from severe misalignment.

\paragraph{Online signal acquisition
(top-5, no extra RPC, overlapped with post-processing).}

In deployment between edge and cloud,
the target distribution is computed on the cloud verifier.
CGSpec obtains $\tilde{p}_t(\cdot\mid x_t)$
without any additional verifier forward pass or extra RPC round
by piggybacking on the standard verification response.
To minimize communication overhead,
we do not transmit distributions for all verified positions.
Instead, the verifier returns the top-5 token IDs
and their conditional probabilities only for the final verified
position in the current sequence.
At the edge, the draft side matches this with the cached
top-5 logits from the corresponding terminal parent node
used during tree growth.
The client then computes $J_t$ using these aligned top-5 supports.
This computation is highly efficient ($k=5$)
and overlapped with acceptance checking and KV-cache reordering,
ensuring the consistency signal is acquired
outside the critical-path waiting time.

\paragraph{Top-5 truncated distributions.}

Let $\mathcal{V}_5^d(x_t)$ and $\mathcal{V}_5^t(x_t)$
denote the draft and target top-5 token-id sets at prefix $x_t$.
We align supports using the union
$\mathcal{V}_5(x_t)
=\mathcal{V}_5^d(x_t)\cup\mathcal{V}_5^t(x_t)$
and construct truncated distributions with a small $\epsilon$
for missing ids:
\begin{equation}
\tilde{p}_m(v\mid x_t)
\propto
\begin{cases}
p_m(v\mid x_t),
& v\in\mathcal{V}_5^m(x_t),\\
\epsilon,
& v\in\mathcal{V}_5(x_t)\setminus\mathcal{V}_5^m(x_t),
\end{cases}
\label{eq:truncated-distribution}
\end{equation}
for $m\in\{d,t\}$,
followed by normalization over $v\in\mathcal{V}_5(x_t)$.
Throughout, we use $k=5$ in the main system
(consistent with our implementation and analysis);
larger $k$ is possible but increases the response payload.

To make the consistency signal suitable for stable online control,
we first normalize the divergence into a bounded mismatch score
\begin{equation}
\hat{J}_t
=
\min\!\left(1,\frac{J_t}{\ln 2}\right),
\label{eq:normalized-js}
\end{equation}
where $\hat{J}_t\in[0,1]$.
A smaller $\hat{J}_t$ indicates stronger local
draft-target agreement,
while a larger value indicates a more difficult speculative state.
We then map this mismatch to an instantaneous control state
\begin{equation}
S_t^{\mathrm{inst}}
=
\max\!\left(S_{\min},(1-\hat{J}_t)^2\right),
\label{eq:instant-consistency}
\end{equation}
where $S_{\min}$ is the minimum admissible scale.
This nonlinear mapping is deliberately risk sensitive:
once mismatch becomes moderately large,
the control state contracts faster than a linear rule,
thereby preventing the system from over expanding speculation
in unreliable states.

Because one speculative round may be influenced
by transient local fluctuation,
we further smooth the state by exponential moving average:
\begin{equation}
S_t
=
\beta_s S_{t-1}
+
(1-\beta_s)S_t^{\mathrm{inst}},
\label{eq:consistency-ema}
\end{equation}
where $\beta_s\in[0,1)$ is the smoothing coefficient.
In addition, when a speculative round is fully accepted,
we do not immediately restore the control state
to its maximum value.
Instead, we apply a gradual recovery rule
\begin{equation}
S_t
=
\min(1,S_{t-1}+\Delta_{\mathrm{rec}}),
\label{eq:consistency-recovery}
\end{equation}
where $\Delta_{\mathrm{rec}}$ is the recovery step.

These updates provide a bounded and smoothed state
for tree adaptation.
The scheduler observes the resulting token gains,
acceptance ratios, and execution times
rather than using the consistency state directly
in its priority rule.%

\paragraph{Scale variable.}

For clarity and reproducibility,
we use the smoothed consistency state as a direct scale variable
for structural adaptation and denote
$s_t\triangleq S_t\in[S_{\min},1]$.

This $s_t$ is updated by the mismatch, smoothing,
and recovery rules above
and controls the growth envelope of the next draft tree.%

\subsection{Nomination and Competition Guided
Speculative Tree Growth (NC Tree)}
\label{sec:NCTree}

In edge-cloud deployment,
speculative tree construction must operate
under a limited upload and verification budget.
Therefore, the goal is not merely to expand locally plausible tokens,
but to retain a globally worthwhile set of candidates
for verification in the cloud.
To this end, we formulate tree construction
as a consistency- and uncertainty-guided
nomination and competition process.

\textbf{The key idea is to separate two roles
that prior methods often mix together.}
Inter-model consistency determines the global growth condition
of the tree, namely, how much speculative budget
should be allocated to depth, width, active paths,
and total retained nodes.
Draft-side uncertainty, by contrast,
is used only for local token nomination under each parent.
The nominated children are then merged into a global candidate pool
and compete again under the total tree budget.
In this way, token exploration is no longer
a simple confidence-based pruning or continuation rule,
but a structured \emph{nomination and competition} process.
Consistency determines how the tree is allowed to grow,
while uncertainty determines which local tokens
are worth proposing within that growth condition.

Formally, let the smoothed consistency state at round $t$ be $S_t$.
We use $S_t$ to determine the structural limits of the current tree,
\begin{equation}
(K_t,L_t,W_t,B_t),
\label{eq:growth-envelope}
\end{equation}
where $K_t$ is the maximum number of active parent paths,
$L_t$ is the maximum number of growth iterations,
$W_t$ is the maximum local branch width per parent 
and $B_t$ is the total node budget.
These variables define the growth envelope of the current round.
A larger $S_t$ corresponds to a more consistent and unfavorable state
and, therefore, permits a deeper and narrower tree,
which emphasizes longitudinal progress.
A smaller $S_t$ indicates a less favorable state
and leads to a shallower and wider tree,
 highlighting horizontal coverage.
Thus, consistency does not rank tokens directly;
it determines the structural condition
in which token growth may proceed.

\paragraph{Explicit mapping
$s_t\rightarrow(K_t,L_t,W_t,B_t)$.}

To make the method fully reproducible,
we provide the explicit control law that maps the scale $s_t$
to the four structural limits.
Let $(K_{\min},K_{\max})$, $(W_{\min},W_{\max})$,
$(B_{\min},B_{\max})$, and $(L_{\min},L_{\max})$
be integer bounds.
We realize the behavior of `` deep/narrow vs.\ shallow/wide 
by mapping depth/budget in the same direction as $s_t$
and mapping width/active paths in the opposite direction:
\begin{align}
L_t
&=
\mathrm{clip}_{[L_{\min},L_{\max}]}
\!\left(
\left\lfloor
L_{\min}+s_t(L_{\max}-L_{\min})
\right\rceil
\right),
\label{eq:depth-map}\\
B_t
&=
\mathrm{clip}_{[B_{\min},B_{\max}]}
\!\left(
\left\lfloor
B_{\min}+s_t(B_{\max}-B_{\min})
\right\rceil
\right),
\label{eq:budget-map}\\
K_t
&=
\mathrm{clip}_{[K_{\min},K_{\max}]}
\!\left(
\left\lfloor
K_{\min}+(1-s_t)(K_{\max}-K_{\min})
\right\rceil
\right),
\label{eq:parent-map}\\
W_t
&=
\mathrm{clip}_{[W_{\min},W_{\max}]}
\!\left(
\left\lfloor
W_{\min}+(1-s_t)(W_{\max}-W_{\min})
\right\rceil
\right).
\label{eq:width-map}
\end{align}

Here, $\lfloor\cdot\rceil$ denotes rounding to the nearest integer,
and $\mathrm{clip}_{[a,b]}$ clamps its argument
to the interval $[a,b]$.%

Given these structural limits,
the speculative tree is grown progressively
in a layer-wise manner.
Let $\mathcal{T}_t^{(\ell)}$ denote the retained tree
after the $\ell$-th growth step.
At each step, the algorithm first identifies
all expandable nodes and selects at most $K_t$
active parents according to accumulated path scores:
\begin{equation}
\mathcal{P}_t^{(\ell)}
=
\mathrm{TopK}_{K_t}
\bigl(
\{\text{candidate nodes}\},
\text{path score}
\bigr).
\label{eq:active-parents}
\end{equation}

For each selected parent
$p\in\mathcal{P}_t^{(\ell)}$,
the draft model performs one forward prediction
and nominates at most $W_t$ local child candidates:
\begin{equation}
\mathcal{C}(p)
=
\mathrm{TopK}_{W_t}
\bigl(p_d(\cdot\mid p)\bigr).
\label{eq:local-nomination}
\end{equation}

Here, the uncertainty of the draft enters the method.
Its role is not to determine the global tree shape,
but to rank local continuations under each parent.
Hence, the confidence in the draft remains fully utilized,
but only for local nomination
rather than for global structural control.

Each nominated child $c\in\mathcal{C}(p)$
is assigned a cumulative path score
\begin{equation}
\mathrm{score}(c)
=
\mathrm{score}(p)
+
\log\delta
+
\log p_d(c\mid p),
\label{eq:path-score}
\end{equation}
and all nominated children are merged into
a global candidate pool:
\begin{equation}
\mathcal{G}_t^{(\ell)}
=
\bigcup_{p\in\mathcal{P}_t^{(\ell)}}\mathcal{C}(p).
\label{eq:candidate-pool}
\end{equation}

The tree is then globally pruned
under the total node budget $B_t$:
\begin{equation}
\mathcal{T}_t^{(\ell+1)}
=
\mathrm{TopB}_{B_t}
\bigl(
\mathcal{T}_t^{(\ell)}
\cup
\mathcal{G}_t^{(\ell)},
\mathrm{score}
\bigr).
\label{eq:global-competition}
\end{equation}

This global competition step is the critical difference
from the purely local token expansion.
A child is not retained simply because it is locally plausible
under its own parent;
it must also justify its use of the shared token budget
against candidates proposed by other parents.
Consequently, a parent may nominate local children,
yet all of them can still be removed
if they are globally dominated.
This global selection allows candidates from different parent nodes to compete for the same node budget, rather than allocating expansion opportunities uniformly across all active parents.

\paragraph{Depth discount $\delta$.}

The factor $\delta\in(0,1)$ applies a fixed penalty
to each additional depth level.
A node at depth $d$ receives a score equal to its log path
probability plus $d\log\delta$.
This penalty favors shallower candidates
when comparing nodes across depths.
Candidates at the same depth receive the same penalty,
so their relative order is determined by path probability.

The depth limit $L_t$ determines how many times
this growth procedure can be repeated
before the tree is submitted for verification.
Taken together, $(K_t,L_t,W_t,B_t)$
define a growth envelope conditioned by consistency,
and the layer-wise nomination and competition procedure
materializes that envelope into a real speculative token tree.
The consistency state directly determines which parent paths can be expanded and how the resulting candidates compete for the available node budget.
Compared with confidence-aware methods,
we do not use draft uncertainty
to determine the global growth condition;
compared to optimization-based tree search methods,
we do not search for a separate offline optimum tree.
Instead, we realize adaptive speculation
as an online tree growth process
that jointly respects local uncertainty
and global budget efficiency.

The resulting tree determines both candidate coverage
and the drafting, upload, and verification costs of each round.
Its performance is therefore evaluated through actual output
progress together with the complete decoding cost.%

\begingroup

\subsection{Overlap-Aware Verification Scheduling}
\label{Sec:CAS}

\begin{figure}[t]
    \centering
    \includegraphics[width=\columnwidth]{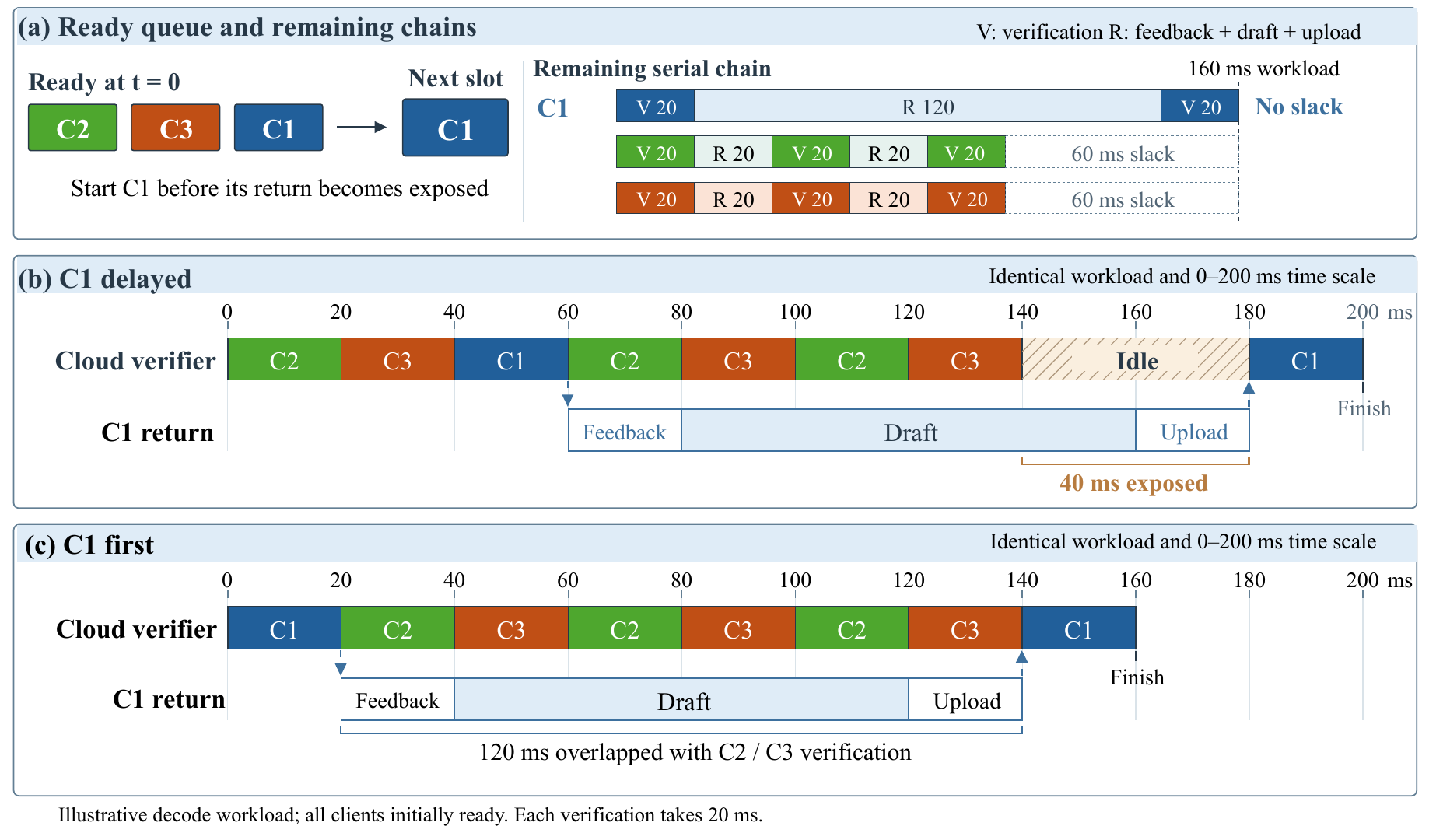}
    \caption{
    Overlap-aware scheduling example.
    (a) Ready clients and remaining serial chains;
    each verification takes 20\,ms.
    (b) Delaying $C_1$ exposes 40\,ms of its return.
    (c) Serving $C_1$ first overlaps its 120\,ms return
    with $C_2$/$C_3$ verification,
    reducing completion time from 200 to 160\,ms.
    Return time includes feedback, drafting, and upload.
    These are illustrative timelines, not measured results.%
    }
    \label{fig:scheduling-overlap}
\end{figure}

Verification order affects both current waiting time
and the time at which each client starts its next
draft feedback.
Figure~\ref{fig:scheduling-overlap} illustrates this dependency.
If the first verification of $C_1$ is delayed,
its long return remains unfinished after $C_2$ and $C_3$
complete their work, exposing 40 \, ms of server idle time.
Serving $C_1$ first allows the other clients' verification work
to cover its entire 120 \,ms return,
reducing completion time for the same workload
from 200 to 160 \,ms.
Scheduling therefore considers each client's remaining
serial chain together with the verification work
available to overlap it.

Because the final output length is unknown before termination,
we estimate the residual output from completed requests.
Let $y_i$ be the decode tokens emitted by the
current request and $\mathcal D_t$ be the multiset of request lengths
completed before time $t$.
With $\mathcal D_t^+(y)=\{d\in\mathcal D_t:d>y\}$,
the regularized mean residual life (MRL) estimate is
\begin{equation}
\widehat r(y)
=
\frac{
\displaystyle
\sum_{d\in\mathcal D_t^+(y)}(d-y)
+\lambda(M-y)/2
}{
|\mathcal D_t^+(y)|+\lambda
},
\label{eq:mrl}
\end{equation}
for $0\le y<M$,
where $M$ is the generation cap
and $\lambda=1$ provides a cold start prior.
For $q_i$ declared requests that have not started,
we use the regularized historical mean $\widehat\mu_t$.
The estimated remaining verification rounds are
\begin{equation}
k_i
=
\frac{
o_i\widehat r(y_i)+q_i\widehat\mu_t
}{
\max(\bar g_i,g_{\min})
},
\label{eq:remaining-rounds}
\end{equation}
where $o_i$ indicates an open request
and $\bar g_i$ is the cumulative actual decode output
divided by the number of decode rounds.
The termination of the request clears the current-request term,
and completed sessions have zero remaining work.
This estimate uses the observed generation progress
without treating the generation cap
as the true output length.

The scheduler maintains the mean
verification-slot duration $S_i$ of each client and the return time $R_i$.
For a ready client, its estimated remaining serial chain
and overlap slack are
\begin{align}
\tau_i
&=
k_iS_i+[k_i-1]_+R_i,
\label{eq:serial-chain}\\
\Delta_i
&=
\sum_{j\in\mathcal A(t)}k_jS_j-\tau_i,
\label{eq:overlap-slack}
\end{align}
where $[x]_+=\max(0,x)$
and $\mathcal A(t)$ include all unfinished clients.
The total of the remaining verification work provides
a reference horizon for covering return intervals.
The smaller $\Delta_i$ indicates less room to postpone the client $i$
without exposing its return.
In the example, $C_1$ has zero slack,
while $C_2$ and $C_3$ each have 60 \,ms.
The slack guides online ranking;
realized overlap also depends on subsequent readiness
and estimation error.

At each free verification slot,
the scheduler applies four rules in order.
First, it serves the oldest ready request
if its wait reaches the protection threshold.
Otherwise, it serves the first ready prefill request.
If no prefill is ready, it selects
the smallest slack ready client satisfying
$\Delta_i\le S_{\mathrm{ref}}$.
If none satisfies this condition, it selects
\begin{equation}
i^*
=
\arg\max_{i\in\mathcal Q(t)}
\frac{a_i}{\max(\epsilon_s,k_iS_i)},
\label{eq:scheduling-utility}
\end{equation}
where $a_i$ is the ratio of the accepted token to the draft node,
$S_{\mathrm{ref}}$ is a reference verification duration,
and $\epsilon_s$ is a numerical floor.
The ties are resolved by a longer waiting time
and then a queue order.
These rules protect waiting requests,
initialize new requests,
start critical return paths early,
and otherwise favor clients with the less remaining
verification work and a higher accepted-path ratio.
Scheduling changes the service order of independent requests
while preserving their target verification
and acceptance rules.

\endgroup

\section{Experimental Setup}
\label{sec:experimental-setup}

\subsection{Experimental Platform and Network Conditions}

\paragraph{Experimental Platform}

The verification servers use AMD Instinct MI250, and Clients are NVIDIA A40s.
The Qwen3-14B and Vicuna-7B target models use one GPU, while Qwen3-32B uses two GPUs with the split KVCache implementation.
Each client process maintains its own draft model state,
generation context, and candidate tree, and submits candidates
to the shared verifier.
The main experiments use $N=8$ and $N=16$ concurrent clients.
The maximum server verification batch size is set to 1
across collaborative methods.

\paragraph{Computation and Communication}

The platform combines GPU model execution with controlled
network emulation based on SSH.
Model forward passes execute on GPUs, while the communication
module injects waiting time according to message size,
uplink and downlink bandwidth, propagation delay, and jitter.
These delays occur during execution and affect verification
request arrivals, server queueing, and the start of subsequent
client drafting.
The measurement window therefore includes model execution,
communication waiting, and their overlap across clients.

\paragraph{Network Conditions}
To emulate realistic edge-cloud deployment conditions,
we evaluate under three representative network profiles
injected through the simulator:
a 4G-like profile with 70 \,ms Round Trip Time (RTT),
10 \,ms jitter, and asymmetric 3/10 \,Mbps
uplink/downlink bandwidth;
a 5G-like profile with 45 \,ms RTT,
3 \,ms jitter, and 30/200 \,Mbps
uplink/downlink bandwidth;
and an edge weak 4G profile with 100 \,ms RTT,
10 \,ms jitter, and 2/10 \,Mbps
uplink/downlink bandwidth.
These settings are based on previous research~\cite{
ghoshal2023performance,lim2023empirical,xu2021cloud}.

The main experiments use a heterogeneous mixture of these
profiles.
The numbers of clients assigned to the 5G-like, 4G-like,
and weak profiles are $(3,2,3)$ for $N=8$ and $(5,6,5)$
for $N=16$.
Additional latency analysis experiments assign the same
5G-like, 4G-like, or weak profile to all clients to examine
how network conditions affect stage costs.

\subsection{Models and Datasets}

\paragraph{Models}

We evaluate four target-draft model pairs:
G1, Qwen3-14B/Qwen3-0.6B;
G2, Qwen3-14B/Qwen3-1.7B;
G3, Vicuna-7B-v1.3/Vicuna-160M;
and G4, Qwen3-32B/Qwen3-1.7B.
G1 and G2 vary draft-model capacity under the same target,
while G2 and G4 vary target-model scale under the same draft
model~\cite{yang2025qwen3}.
G3 introduces an additional model family.
These configurations examine system behavior as draft
capability and target verification cost change.

\paragraph{Datasets}
We use three datasets with complementary workload characteristics:
\textbf{C4}~\cite{JMLR:v21:20-074}
for general domain system efficiency evaluation,
\textbf{WikiText-2}~\cite{merity2016pointer}
for the continuous analysis from long form generation
and verifier amortization analysis,
and \textbf{SpecBench}~\cite{xia2024survey}
for heterogeneous interactive tasks
that stress robustness across diverse request patterns.
In each set of experiments, all clients divide this data equally among themselves.

\subsection{Compared Methods and Common Execution Conditions}

\begingroup

We compare CGSpec with four baselines:

\begin{itemize}
    \item
    \textbf{Target-AR}
    performs autoregressive generation with the target model
    on the server, providing a non-speculative reference.

    \item
    \textbf{EAGLE-3}~\cite{li2026eagle}
    serves as the single server speculative decoding baseline
    for all four model groups.
    Both drafting and verification are executed on the server.

    \item
    \textbf{SLED}~\cite{li2025sled}
    performs client chain drafting and uses server
    FIFO dispatch.
    Its verification batch size is set to one in the
    main experiments.

    \item
    \textbf{SpecEdge}~\cite{park2026specedge}
    uses its client tree generation strategy with
    server FIFO dispatch.
\end{itemize}

\paragraph{Common Conditions}
All methods use the same target model, request plan,
client count, generation temperature, and output-token cap.
SLED, SpecEdge, and CGSpec additionally share the same model pair, network conditions,
serving harness, and server verification batch size.
Comparisons among these collaborative methods evaluate
candidate generation and verification scheduling policies
under these common conditions.

\paragraph{Baseline Roles}
Target-AR and EAGLE-3 provide cloud autoregressive
and cloud speculative references.
SLED and SpecEdge provide collaborative decoding baselines.
We report throughput for all methods and summarize aggregate
relative improvements using SpecEdge as the primary
collaborative reference.


\endgroup

\subsection{Workload and Measurement Protocol}

\begingroup

\paragraph{Request Plans}
The main experiments use temperature zero and a maximum
decode budget of 128 tokens per request.
Each measured run contains a total of 128 requests,
randomly assigned to all $N$ clients.
Each request is assigned to one client.
For a given seed and client count, the request set and its
assignment are reused across methods.
The total measured request count is 128 for both $N=8$
and $N=16$.

\paragraph{Warmup and Measurement Window}
Each client first executes one warmup request.
After all clients finish warmup, a synchronization barrier
starts the common measurement window.
The window ends when the last client completes its assigned
requests.
Warmup requests are excluded from both the measured output
and the measurement window.
Token counts use actual generated output, including
differences caused by early termination.

\paragraph{Repeated Runs and Paired Statistics}
Each configuration is evaluated
with 5 seeds, 42--46.
We report the mean and standard deviation of
the sample in these runs.
Relative improvements are computed by pairing methods
within the same model, dataset, client count, and seed.
To summarize aggregate improvement, we first calculate result for each
configuration within a seed, average these ratios across
configurations, and then average across seeds.

\paragraph{Workload Coverage}
The main experiments cover four pair of models, three datasets 
and two client counts, yielding 24 configurations. 
The scheduling scalability experiment uses G1 and SpecBench
under mixed networks, varying the client count from 2 to 30.
Client counts, request counts, and repetitions are reported
with the corresponding experimental results.

\endgroup

\subsection{Evaluation Metrics}

\begingroup

\paragraph{System Throughput}
System throughput is the total number of actual output
tokens between clients divided by the common measurement
window.
 Measures aggregate generation progress supported by
the shared verifier.

\paragraph{Client Throughput}
For each client, we divide its actual output token count
by the interval from the start of the common measurement to the
completion of all assigned requests.
Client throughput is the mean of these per-client rates,
capturing individual generation progress under shared
verification.

\paragraph{Request Latency and Stage Costs}
End-to-end request latency measures the interval from
request start to delivery of the complete response.
The time breakdown distinguishes local drafting,
communication, server queueing, and verification.
Server queueing spans request arrival to service start,
while verification covers the full server service slot.
For the stage breakdown plots, we multiply each stage's
mean per-round duration by the number of rounds per request
and aggregate the resulting costs across datasets and seeds.

\paragraph{Scheduling Wait}
The scheduling comparison also reports client RPC wait,
measuring the time a client waits for a verification call
to return.
This metric captures waiting during the verification
interaction and is recorded separately from server-side
queueing before service begins.

\paragraph{Speculative Efficiency}
We report the number of accepted draft tokens per
verification round and the ratio of accepted draft
tokens to submitted draft nodes.
These metrics measure the accepted length and candidate
utilization, respectively.
Actual emitted tokens may additionally include a
target token and are used for throughput
calculation and remaining-work estimation.

\endgroup

\subsection{Component Evaluation}
\label{sec:component-evaluation}

\begingroup

\paragraph{Consistency Signal and Budget Control}
We separately examine the relationship between consistency
and accepted length, and the performance effect of using
consistency for online control.
The signal analysis fixes the candidate budget at 64 nodes
per round and disables consistency-driven budget adaptation.
It records the relationship between Top-$k$ JS divergence
and accepted length.
The end-to-end ablation compares a fixed tree,
confidence-based pruning, acceptance-rate guidance,
JS-based budget control alone, and full CGSpec to evaluate
how these control strategies affect throughput.

\paragraph{Candidate-Tree Construction}
In the fixed-budget experiment, probability-driven tree
construction and nomination and competition use the same
per-round node budget, with consistency-driven budget
adaptation disabled.
This comparison examines accepted progress under a
common candidate budget.
The tree-control ablation uses G1 and G3 on SpecBench
under mixed networks, with $N=8$ and five seeds.
All tree control variants retain the same proposed
verification scheduler, focusing the comparison on
client-side candidate generation and budget control.

\paragraph{Verification Scheduling}
We compare FIFO, LAS \cite{rai2003analysis}, LWF \cite{edmonds2005maiden}, SRPT \cite{smith1978new} and CGSpec to examine
priorities based on arrival order, service achieved,
waiting time, estimated serving cost and overlap-aware
decisions.
LAS uses accumulated token progress, LWF uses server 
queueing time, and SRPT uses the estimated serving cost
in our implementation.
The comparison uses G1 and SpecBench under mixed networks reports system throughput, mean client
throughput, and client RPC wait.

\endgroup

\begin{table*}[t]
\centering
\caption{
System and client throughput under mixed network conditions.
Values are mean $\pm$ sample standard deviation over five seeds,
in tokens/s. Sys and Cli denote system throughput and mean client
throughput, respectively. Bold values indicate the highest mean
within each configuration.
}
\label{tab:main_results}

\begingroup
 
\setlength{\tabcolsep}{3pt}
\renewcommand{\arraystretch}{1.12}

\resizebox{\textwidth}{!}{%
\begin{tabular}{ll*{5}{rr}}
\toprule
Model pair & Dataset
& \multicolumn{2}{c}{Target-AR}
& \multicolumn{2}{c}{EAGLE 3}
& \multicolumn{2}{c}{SLED}
& \multicolumn{2}{c}{SpecEdge}
& \multicolumn{2}{c}{CGSpec} \\
\cmidrule(lr){3-4}
\cmidrule(lr){5-6}
\cmidrule(lr){7-8}
\cmidrule(lr){9-10}
\cmidrule(lr){11-12}
& &
Sys $\uparrow$ & Cli $\uparrow$ &
Sys $\uparrow$ & Cli $\uparrow$ &
Sys $\uparrow$ & Cli $\uparrow$ &
Sys $\uparrow$ & Cli $\uparrow$ &
Sys $\uparrow$ & Cli $\uparrow$ \\

\midrule
\multicolumn{12}{c}{$N=8$} \\
\midrule

\multirow{3}{*}{\shortstack[l]{G1: Qwen3\\14B/0.6B}}
& SpecBench
& $19.8\pm0.1$ & $2.59\pm0.01$
& $36.8\pm0.2$ & $4.82\pm0.02$
& $35.0\pm6.7$ & $4.71\pm0.96$
& $54.7\pm1.8$ & $7.65\pm0.06$
& $\mathbf{60.7\pm1.6}$ & $\mathbf{8.43\pm0.05}$ \\
& C4
& $19.8\pm0.1$ & $2.60\pm0.01$
& $30.1\pm0.9$ & $3.95\pm0.11$
& $35.6\pm0.8$ & $4.70\pm0.12$
& $53.6\pm1.7$ & $7.46\pm0.14$
& $\mathbf{58.5\pm2.0}$ & $\mathbf{7.98\pm0.16}$ \\
& WikiText-2
& $19.4\pm0.2$ & $2.54\pm0.03$
& $26.1\pm0.5$ & $3.43\pm0.06$
& $31.6\pm5.3$ & $4.16\pm0.69$
& $50.3\pm0.9$ & $6.99\pm0.10$
& $\mathbf{54.7\pm1.4}$ & $\mathbf{7.34\pm0.15}$ \\

\addlinespace[2pt]
\multirow{3}{*}{\shortstack[l]{G2: Qwen3\\14B/1.7B}}
& SpecBench
& $19.8\pm0.1$ & $2.60\pm0.01$
& $36.8\pm0.4$ & $4.82\pm0.03$
& $43.9\pm0.9$ & $5.87\pm0.08$
& $52.8\pm11.5$ & $7.46\pm1.22$
& $\mathbf{65.5\pm1.7}$ & $\mathbf{8.99\pm0.06}$ \\
& C4
& $19.8\pm0.1$ & $2.59\pm0.01$
& $30.2\pm0.9$ & $3.96\pm0.11$
& $39.5\pm1.1$ & $5.23\pm0.13$
& $58.3\pm1.0$ & $8.03\pm0.12$
& $\mathbf{65.5\pm3.4}$ & $\mathbf{8.94\pm0.18}$ \\
& WikiText-2
& $19.4\pm0.2$ & $2.54\pm0.02$
& $26.2\pm0.5$ & $3.43\pm0.08$
& $38.6\pm1.0$ & $5.08\pm0.11$
& $55.7\pm0.8$ & $7.59\pm0.10$
& $\mathbf{62.1\pm1.3}$ & $\mathbf{8.19\pm0.12}$ \\

\addlinespace[2pt]
\multirow{3}{*}{\shortstack[l]{G3: Vicuna\\7B/160M}}
& SpecBench
& $35.5\pm0.2$ & $4.64\pm0.01$
& $64.9\pm0.5$ & $8.49\pm0.07$
& $48.7\pm1.7$ & $6.96\pm0.08$
& $59.3\pm2.3$ & $10.57\pm0.19$
& $\mathbf{65.7\pm3.4}$ & $\mathbf{10.95\pm0.17}$ \\
& C4
& $35.9\pm0.4$ & $4.70\pm0.06$
& $62.6\pm0.5$ & $8.21\pm0.05$
& $52.5\pm0.7$ & $7.24\pm0.19$
& $67.2\pm2.1$ & $11.27\pm0.31$
& $\mathbf{73.6\pm2.9}$ & $\mathbf{11.46\pm0.23}$ \\
& WikiText-2
& $33.6\pm0.9$ & $4.40\pm0.12$
& $49.8\pm1.3$ & $6.54\pm0.20$
& $45.0\pm0.7$ & $6.12\pm0.16$
& $55.4\pm2.9$ & $9.13\pm0.44$
& $\mathbf{61.7\pm3.3}$ & $\mathbf{9.47\pm0.30}$ \\

\addlinespace[2pt]
\multirow{3}{*}{\shortstack[l]{G4: Qwen3\\32B/1.7B}}
& SpecBench
& $9.5\pm0.0$ & $1.24\pm0.00$
& $19.1\pm0.2$ & $2.50\pm0.01$
& $21.6\pm0.2$ & $2.80\pm0.02$
& $29.4\pm0.4$ & $3.84\pm0.02$
& $\mathbf{32.6\pm0.3}$ & $\mathbf{5.03\pm0.18}$ \\
& C4
& $9.5\pm0.0$ & $1.24\pm0.01$
& $14.3\pm0.3$ & $1.87\pm0.04$
& $18.6\pm0.2$ & $2.41\pm0.04$
& $28.6\pm1.7$ & $3.69\pm0.22$
& $\mathbf{31.4\pm1.7}$ & $\mathbf{4.83\pm0.39}$ \\
& WikiText-2
& $9.2\pm0.1$ & $1.21\pm0.01$
& $12.4\pm0.3$ & $1.63\pm0.04$
& $17.8\pm0.9$ & $2.30\pm0.11$
& $27.1\pm0.1$ & $3.50\pm0.04$
& $\mathbf{29.9\pm0.2}$ & $\mathbf{4.60\pm0.08}$ \\

\midrule
\multicolumn{12}{c}{$N=16$} \\
\midrule

\multirow{3}{*}{\shortstack[l]{G1: Qwen3\\14B/0.6B}}
& SpecBench
& $19.7\pm0.0$ & $1.31\pm0.02$
& $35.5\pm0.1$ & $2.37\pm0.03$
& $38.0\pm3.7$ & $3.06\pm0.41$
& $58.4\pm0.5$ & $4.64\pm0.03$
& $\mathbf{65.2\pm1.3}$ & $\mathbf{6.06\pm0.13}$ \\
& C4
& $20.0\pm0.1$ & $1.33\pm0.01$
& $30.0\pm0.4$ & $2.00\pm0.03$
& $35.2\pm1.2$ & $2.32\pm0.09$
& $57.2\pm1.4$ & $3.76\pm0.07$
& $\mathbf{61.8\pm1.0}$ & $\mathbf{5.08\pm0.12}$ \\
& WikiText-2
& $19.3\pm0.2$ & $1.28\pm0.01$
& $25.7\pm0.2$ & $1.71\pm0.02$
& $34.1\pm0.5$ & $2.21\pm0.04$
& $52.8\pm0.2$ & $3.46\pm0.03$
& $\mathbf{56.7\pm0.7}$ & $\mathbf{4.54\pm0.10}$ \\

\addlinespace[2pt]
\multirow{3}{*}{\shortstack[l]{G2: Qwen3\\14B/1.7B}}
& SpecBench
& $19.7\pm0.0$ & $1.31\pm0.01$
& $35.7\pm0.2$ & $2.38\pm0.03$
& $45.1\pm0.6$ & $3.69\pm0.04$
& $60.4\pm1.3$ & $4.84\pm0.08$
& $\mathbf{70.7\pm0.6}$ & $\mathbf{6.60\pm0.09}$ \\
& C4
& $19.9\pm0.1$ & $1.32\pm0.00$
& $30.0\pm0.4$ & $2.00\pm0.03$
& $39.5\pm0.6$ & $2.60\pm0.03$
& $62.3\pm1.4$ & $4.04\pm0.08$
& $\mathbf{71.0\pm0.8}$ & $\mathbf{5.69\pm0.05}$ \\
& WikiText-2
& $19.3\pm0.2$ & $1.29\pm0.01$
& $25.6\pm0.3$ & $1.70\pm0.02$
& $38.6\pm0.4$ & $2.50\pm0.02$
& $57.7\pm0.7$ & $3.75\pm0.03$
& $\mathbf{64.4\pm0.4}$ & $\mathbf{5.14\pm0.11}$ \\

\addlinespace[2pt]
\multirow{3}{*}{\shortstack[l]{G3: Vicuna\\7B/160M}}
& SpecBench
& $35.6\pm0.4$ & $2.36\pm0.03$
& $63.2\pm1.0$ & $4.24\pm0.08$
& $49.7\pm0.7$ & $3.66\pm0.03$
& $73.0\pm3.0$ & $5.42\pm0.05$
& $\mathbf{74.5\pm2.6}$ & $\mathbf{6.17\pm0.16}$ \\
& C4
& $35.9\pm0.3$ & $2.40\pm0.03$
& $63.6\pm0.7$ & $4.24\pm0.04$
& $53.2\pm1.1$ & $3.64\pm0.04$
& $80.2\pm2.6$ & $5.79\pm0.10$
& $\mathbf{87.0\pm3.6}$ & $\mathbf{7.18\pm0.34}$ \\
& WikiText-2
& $32.1\pm0.8$ & $2.13\pm0.05$
& $48.0\pm1.9$ & $3.20\pm0.14$
& $45.7\pm0.8$ & $3.01\pm0.03$
& $66.7\pm1.2$ & $4.70\pm0.06$
& $\mathbf{68.2\pm1.3}$ & $\mathbf{5.60\pm0.22}$ \\

\addlinespace[2pt]
\multirow{3}{*}{\shortstack[l]{G4: Qwen3\\32B/1.7B}}
& SpecBench
& $9.4\pm0.1$ & $0.62\pm0.01$
& $18.7\pm0.0$ & $1.24\pm0.02$
& $21.8\pm0.1$ & $1.78\pm0.02$
& $29.1\pm1.3$ & $2.30\pm0.08$
& $\mathbf{33.3\pm0.4}$ & $\mathbf{3.33\pm0.07}$ \\
& C4
& $9.5\pm0.1$ & $0.63\pm0.01$
& $14.2\pm0.1$ & $0.95\pm0.01$
& $18.6\pm0.1$ & $1.20\pm0.01$
& $29.5\pm0.2$ & $1.88\pm0.01$
& $\mathbf{32.7\pm0.1}$ & $\mathbf{2.68\pm0.08}$ \\
& WikiText-2
& $9.2\pm0.1$ & $0.61\pm0.01$
& $12.3\pm0.2$ & $0.82\pm0.01$
& $18.5\pm0.2$ & $1.18\pm0.02$
& $26.7\pm1.3$ & $1.71\pm0.08$
& $\mathbf{30.5\pm0.5}$ & $\mathbf{2.40\pm0.13}$ \\

\bottomrule
\end{tabular}%
}
\endgroup
\end{table*}

\section{Experimental Results}

\subsection{End-to-End Throughput}

\begingroup

Table~\ref{tab:main_results} reports system throughput and
client throughput under mixed network conditions for four model
pairs, three datasets, and two client counts. CGSpec achieves
the highest mean for both metrics in all 24 configurations.
Compared to SpecEdge, CGSpec improves system performance by
12.2\% at $N=8$ and 10.2\% at $N=16$, while improving mean
client performance by 14.0\% and 33.0\%, respectively.
We compute these aggregate gains from the paired throughput ratios
within each configuration and then average
between the configurations and seeds.

The gains vary across model pairs. At $N=16$, the average
improvement in system throughput over SpecEdge is 9.0\% for G1,
14.2\% for G2, 4.3\% for G3, and 13.3\% for G4.
For example, on G2/C4, CGSpec increases system throughput from
62.3 to 71.0 tokens/s and mean client throughput from
4.04 to 5.69 tokens/s. On G4/SpecBench, the corresponding
values increase from 29.1 to 33.3 tokens/s and from
2.30 to 3.33 tokens/s.
The higher aggregate client throughput gain at $N=16$ shows
that the benefit extends to individual client progress under
higher concurrency, even though the system throughput gain
is smaller than at $N=8$.

Compared to SLED, CGSpec improves system throughput by
58.3\% at $N=8$ and 64.6\% at $N=16$.
The corresponding gains over EAGLE-3 are 81.2\% and 95.7\%.
EAGLE-3 provides a single speculative decoding
in the main experiment, with both drafting and verification
performed on the server; the collaborative methods additionally
use client drafting resources.
These comparisons quantify throughput under the evaluated
deployment configurations. Component ablations separately
examine the contributions of tree construction and verification
scheduling.

\endgroup

\subsection{Speculative Efficiency Analysis}
\label{sec:speculative-efficiency}

To further explain where the end-to-end throughput gain comes from, we first analyze the difference in speculative efficiency of our method and baselines. Figure~\ref{fig:efficiency_overall} characterizes speculative behavior using two key metrics: \textit{token acceptance rate} and \textit{mean accepted length}. These two metrics correspond to the two core objectives of our design. A higher acceptance rate means that drafted tokens are more likely to survive target verification, thereby reducing wasted communication and verification time. A longer accepted length means that each cloud-side verification produces more real decoding progress, thus improving verifier amortization.

As shown in Figure~\ref{fig:efficiency_overall}, CGSpec consistently occupies a more favorable upper-right region, indicating that it improves both acceptance rate and the accepted length at the same time rather than trading one against the other. Compared with SpecEdge, our method does not simply obtain a higher acceptance rate by conservatively shrinking the speculative budget. Instead, it suppresses ineffective expansion while still preserving enough high value candidates, so both metrics improve simultaneously. Compared with SLED, CGSpec submits more candidate tokens than a purely chain-based draft, and therefore obtains longer effective accepted strides during verification. The SLED chain structure can often maintain relatively high local acceptance, but once an early token is rejected, later tokens in the chain cannot be preserved, so its accepted length is inherently limited. SpecEdge lies at the opposite extreme: it explores more candidates with a larger tree, but when draft-target consistency is insufficient, that larger tree quickly turns into extra communication cost and verification waste. By contrast, CGSpec first controls the total draft-token budget through consistency-aware sensing, and then controls tree construction through nomination and competition, thereby avoiding both extremes of being overly conservative and overly aggressive.

We further investigate the performance of the task variation by dividing the SpecBench dataset into nine task categories and compare the token acceptance rate and the mean accepted length of CGSpec with the strong baseline SpecEdge in each category shown in Figure~\ref{fig:efficiency_tasks}. CGSpec improves speculative efficiency across most task categories, rather than benefiting only from aggregate performance gains.
Another important observation is that CGSpec exhibits much greater variation (Figure~\ref{fig:efficiency_overall}), driven primarily by differences in the accepted length between tasks (Figure~\ref{fig:efficiency_tasks}). This variation reflects the task behavior of CGSpec. The method expands more aggressively on tasks with higher draft–target consistency, such as Math, Extraction, and Translation, while adopting more conservative speculation on tasks with lower consistency. This confirms that it can dynamically adapt the speculative intensity according to the consistency of draft-target about the task.

\begin{figure*}[htbp]
    \centering
    \includegraphics[width=0.9\textwidth]
    {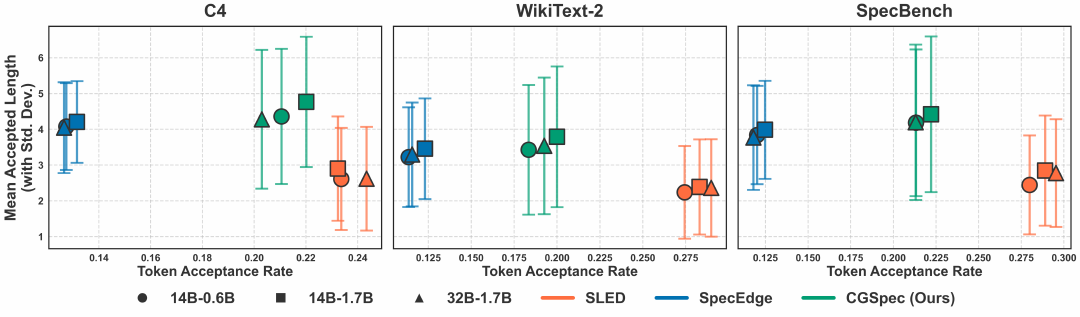}
    \caption{\textbf{Speculative efficiency distribution across datasets for different methods and model pairs.} The x-axis is token acceptance rate, and the y-axis shows the mean accepted length per verification step. Points closer to the upper-right corner indicate simultaneously higher acceptance rate and longer accepted length.}
    \label{fig:efficiency_overall}
\end{figure*}

\begin{figure*}[htbp]
    \centering
    \includegraphics[width=0.9\textwidth]
    {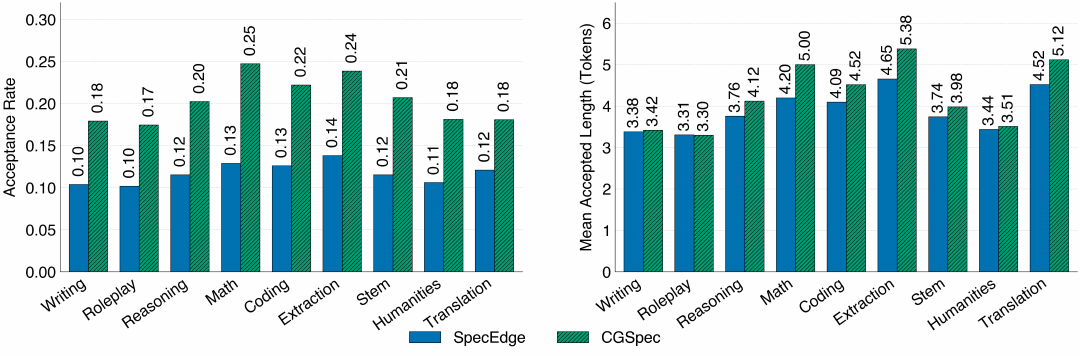}
    \caption{\textbf{Speculative efficiency decomposition across task categories.} The left panel shows token acceptance rate for each task category, and the right panel shows the corresponding mean accepted length. The x-axis denotes task categories, and the y-axis denotes the corresponding metric values. Different colors denote different methods.}
    \label{fig:efficiency_tasks}
\end{figure*}

\subsection{Latency Breakdown}
\label{sec:latency-breakdown}
\label{sec:main-throughput}

\begin{figure}
    \centering
    \includegraphics[width=\columnwidth]
    {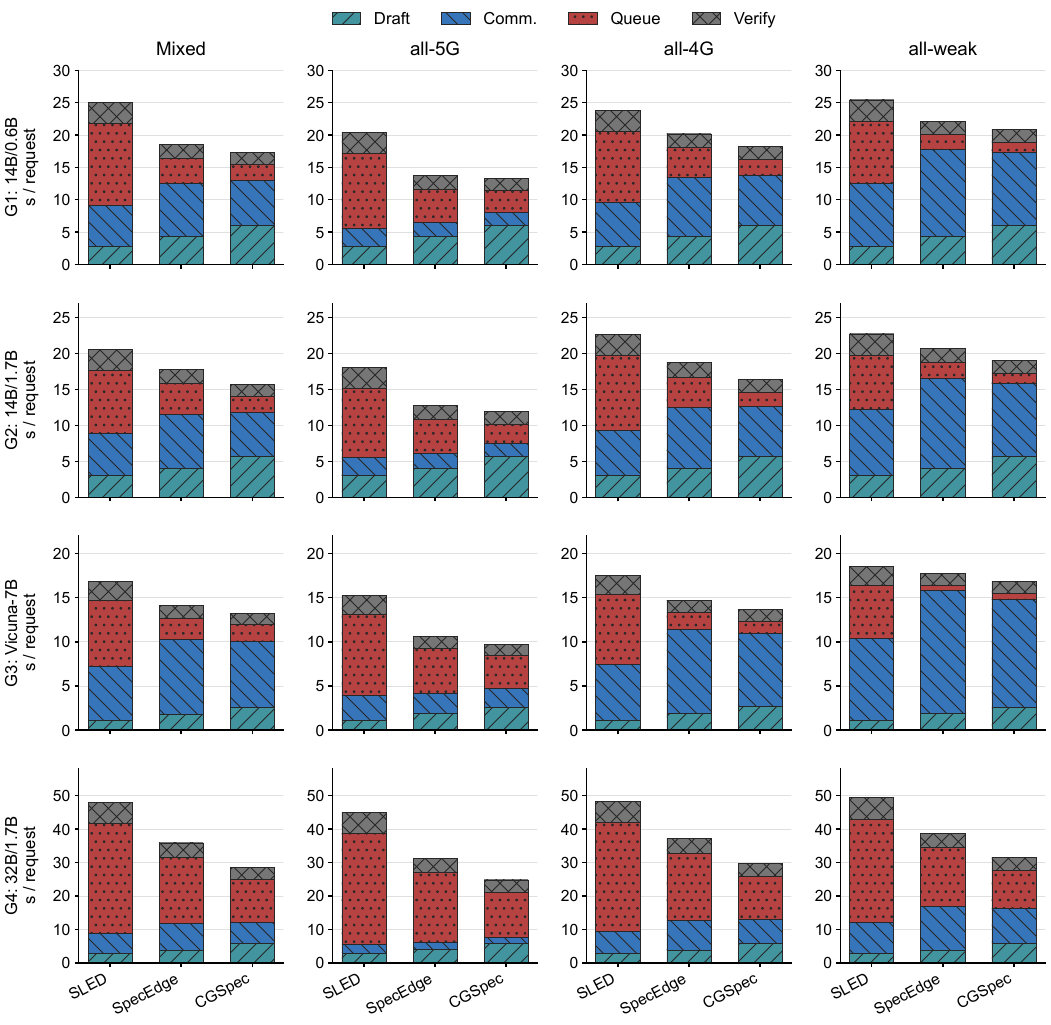}
    \caption{
    Cumulative stage time per request under different networks.
    }
    \label{fig:delay_breakdown}
\end{figure}

\begingroup

Figure~\ref{fig:delay_breakdown} decomposes cumulative
per-request time into drafting, communication, queueing,
and verification at $N=8$.
Rows correspond to the four model pairs, while columns
represent mixed, all-5G, all-4G, and all-weak networks.
For each run, each component is estimated by multiplying
its mean per-round duration by the number of rounds
per request, and the results are averaged over three
datasets and five seeds.
Queueing measures server-side waiting before verification
begins. The stacked bars describe accumulated stage costs.

Across the evaluated configurations, CGSpec spends more
time on drafting than SpecEdge, but less time on
communication, queue, and verification.
For example, under the mixed networks with G1,
drafting time increases from 4.27 to 6.06 \,s per request,
while communication time decreases from 8.19 to 6.89 \,s,
queue time from 3.90 to 2.48 \,s,
and the verification time from 2.14 to 1.94 \,s.
The additional drafting cost is therefore offset by
reductions in the remaining stages.
This pattern shows that reducing local drafting time
alone is insufficient to minimize the accumulated
cost of collaborative generation.

The network conditions determine which components
dominate the breakdown.
In all-5G, communication occupies a relatively small
fraction of the accumulated time, making the differences
in server waiting more visible.
For G1, CGSpec reduces queue time from
5.14 to 3.38 \,s relative to the SpecEdge,
whereas the communication time decreases from
2.24 to 1.96 \,s.
In all-weak, communication becomes the largest
component for G1--G3.
For G1, it reaches 13.48 \,s with SpecEdge and
11.30 \,s with CGSpec, while queueing contributes
2.28 and 1.56 \,s, respectively.
CGSpec reduces both components, but the substantial
communication cost remains in the weaker network.

The model pairs also exhibit different cost profiles.
For G4, queueing remains the largest component across
all four network conditions.
In mixed networks, CGSpec reduces the queue time
from 19.81 to 12.64\,s per request,
while the drafting time increases from 3.88 to 5.69 \,s.
In contrast, for G3 under mixed networks,
communication dominates: CGSpec spends 7.45 \,s
on communication and 1.86 \,s on queueing.
These differences show why the same optimization
can have different effects across model and network
configurations.

The SLED further illustrates the tradeoff between inexpensive
drafting and downstream costs.
For G1 under mixed networks, SLED requires only
2.81 \,s of drafting time per request,
compared to 6.06 \,s for CGSpec.
However, its accumulated queueing and verification
times are 12.60 and 3.29 \,s,
compared to 2.48 and 1.94 \,s for CGSpec.
The breakdown thus identifies lower waiting and
verification costs as important counterparts to
CGSpec's additional drafting work.
Because these comparisons vary both drafting and
scheduling policies, the subsequent ablation study
isolates their individual contributions.

\endgroup

\subsection{Effectiveness of Consistency-Guided Control}
\label{sec:consistency-control}

We evaluate whether JS divergence provides
an informative signal for accepted length under a fixed
speculative budget. This experiment examines the
relationship independently of online consistency-guided
budget adaptation.

Under fixed network and model settings, we record the token-level and step-level top-level JS divergence $k$ and analyze its relationship with the accepted length. In this experiment, JS is used only as an observation signal and does not participate in online control. The fixed-tree baseline is assigned a constant budget of 64 tokens per round and the token selection is purely driven by probability. Candidate parent nodes are first ranked by cumulative draft log-probability, each parent expands its local top-k tokens, and the resulting children are globally pruned by cumulative log-probability under a fixed budget constraint. CGSpec is evaluated under the same token budget without invoking consistency-based control, while still using the nomination and competition procedure to construct the speculative tree.

\begin{figure*}[t]
    \centering
    \subfloat[Correlation vs. Top-$K$ JS Divergence]{
        \includegraphics[width=0.31\textwidth]
        {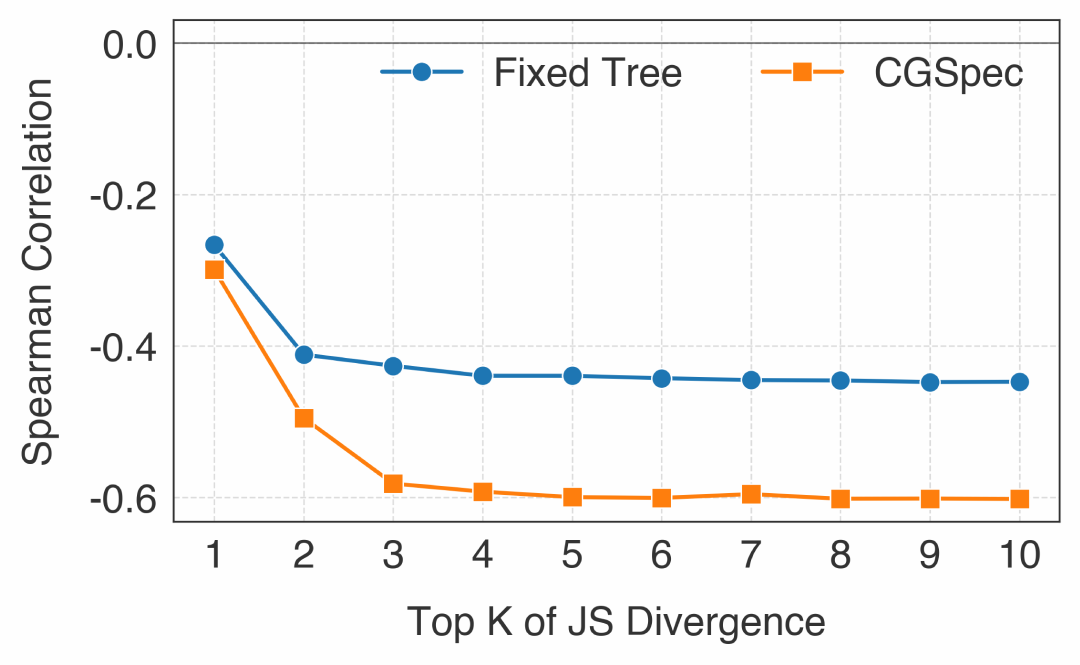}
    }\hfill
    \subfloat[Accepted length vs. Top5 JS (0.6/14B)]{
        \includegraphics[width=0.31\textwidth]
        {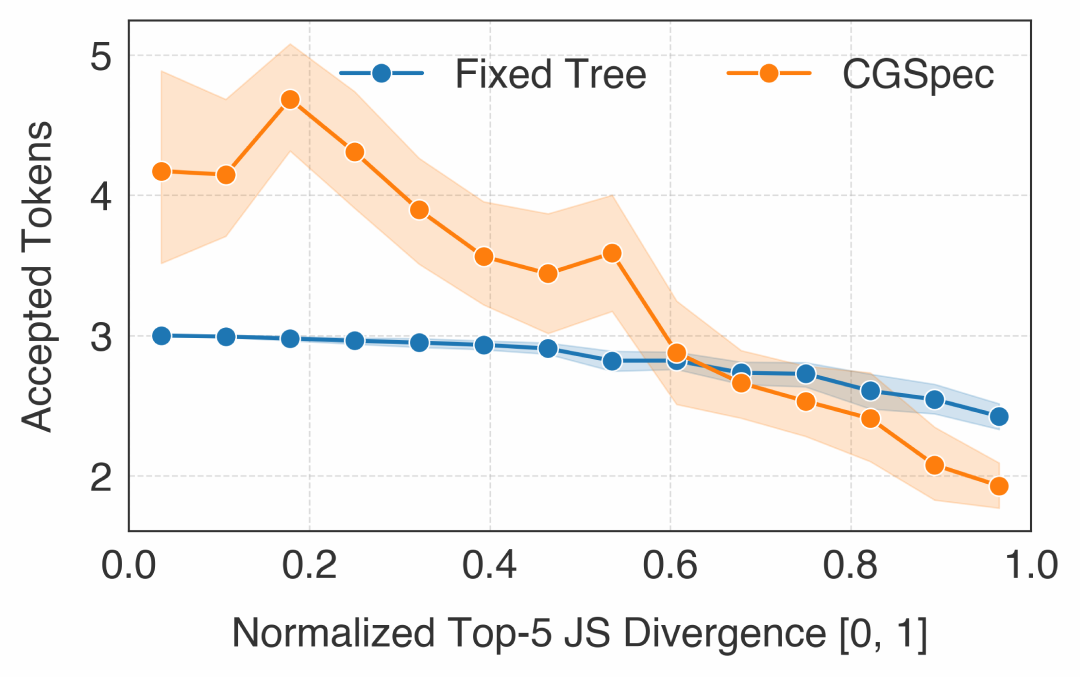}
    }\hfill
    \subfloat[Accepted length vs. Top5 JS (1.7/32B)]{
        \includegraphics[width=0.31\textwidth]
        {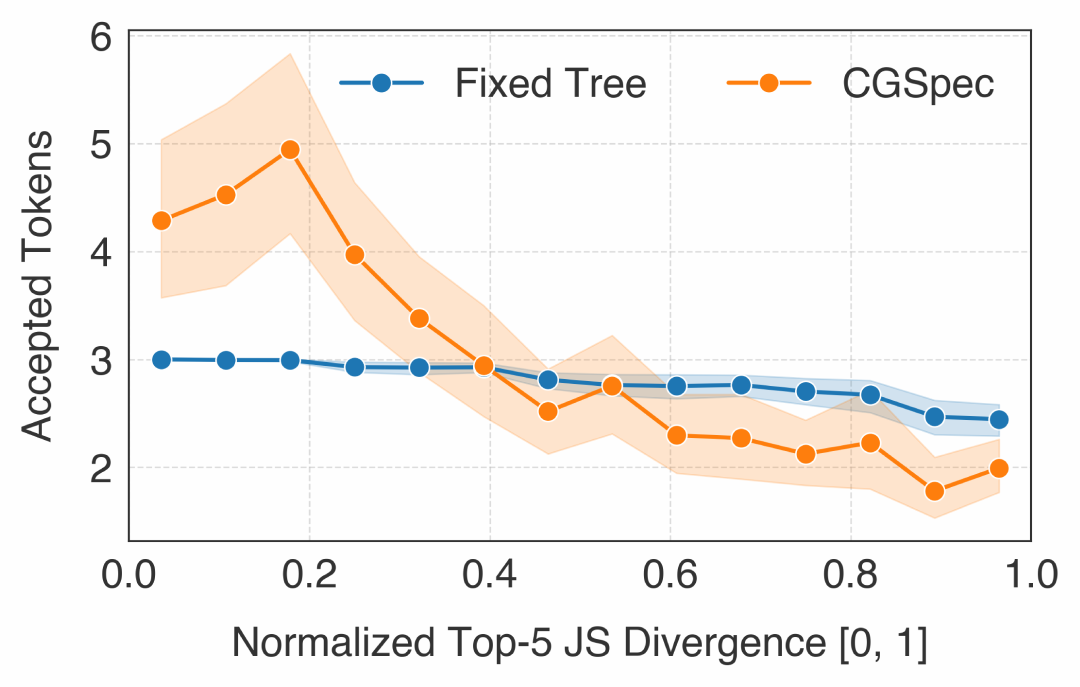}
    }
    \caption{\textbf{Consistency-guided control analysis.}
    \textbf{(a)} Spearman correlation between Top-$k$
    JS divergence and accepted length.
    \textbf{(b)--(c)} Accepted length versus normalized
    Top-5 JS divergence under two model settings.
    Points denote binned means, solid lines denote
    smoothed trends, shaded bands denote 95\%
    confidence intervals.}
    \label{fig:consistency_control}
\end{figure*}

Figure~\ref{fig:consistency_control}(a) shows a negative
spearman correlation between Top-$k$ JS divergence and
accepted length for both tree construction methods.
The correlation strengthens as $k$ increases from 1
to approximately 5 and changes little thereafter,
supporting Top-5 as a compact choice for consistency
sensing.
Figures~\ref{fig:consistency_control}(b)--(c) further
show that accepted length generally decreases as
normalized Top-5 JS divergence increases under both
model settings.
Accepted length decreases as draft–target divergence increases in both model settings.

\begin{figure*}[t]
    \centering
    \includegraphics[width=\textwidth]
    {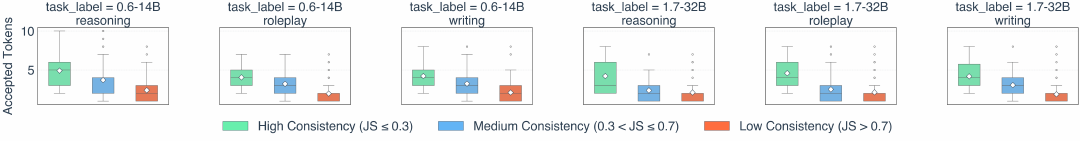}
    \caption{\textbf{Accepted length distribution under
    JS-based state partitioning across different task
    categories.} Each panel corresponds to one task
    type, and the y-axis reports accepted tokens per
    verification step. Colors denote three consistency
    states defined by JS thresholds:
    low-JS (\(JS \leq 0.3\)),
    medium-JS (\(0.3 < JS \leq 0.7\)),
    and high-JS (\(JS > 0.7\)).
    White dots indicate the mean, center lines indicate
    the median, boxes show the interquartile range,
    whiskers represent the non-outlier range, and points
    outside the whiskers denote a small number of outliers.}
    \label{fig:js_states}
\end{figure*}

Figure~\ref{fig:js_states} examines this relationship
within reasoning, roleplay, and writing tasks for
the two model pairs.
Using the original thresholds, we partition observations
into low - JS ($JS\leq0.3$), medium - JS
($0.3<JS\leq0.7$), and high - JS ($JS>0.7$) states.
Mean accepted length follows the ordering
low - JS $>$ medium - JS $>$ high - JS in all six panels.
This supports the use of JS
divergence to distinguish states with different
expected accepted progress.
The ablation study then evaluates the performance
contribution of using this signal for online control.

\subsection{Ablation Study}
\label{sec:ablation}

\begin{table*}[t]
\centering
\caption{
Tree construction ablation on SpecBench under mixed
networks with $N=8$.
Values are mean $\pm$ sample standard deviation over
five seeds, in tokens/s.
All variants retain the proposed verification scheduler.
Full CGSpec results are taken from the corresponding
main experiments.
Bold indicates the highest mean in each column.
}
\label{tab:ablation_study}

\begingroup

\setlength{\tabcolsep}{8pt}

\begin{tabular}{lrrrr}
\toprule
Variant
& \multicolumn{2}{c}{G1: Qwen3-14B/0.6B}
& \multicolumn{2}{c}{G3: Vicuna-7B/160M} \\
\cmidrule(lr){2-3}
\cmidrule(lr){4-5}
& System & Client & System & Client \\
\midrule

Fixed tree, no JS
& $43.58\pm1.14$ & $6.11\pm0.13$
& $55.00\pm1.92$ & $8.90\pm0.26$ \\

Confidence pruning
& $47.60\pm0.79$ & $7.03\pm0.12$
& $53.52\pm2.00$ & $8.40\pm0.48$ \\

Acceptance-rate guidance
& $53.05\pm0.96$ & $7.29\pm0.16$
& $61.14\pm2.74$ & $10.20\pm0.24$ \\

JS budget only
& $52.08\pm1.09$ & $7.32\pm0.27$
& $\mathbf{66.53\pm2.31}$ & $9.98\pm0.43$ \\

CGSpec (full)
& $\mathbf{60.65\pm1.64}$ & $\mathbf{8.43\pm0.05}$
& $65.71\pm3.43$ & $\mathbf{10.95\pm0.17}$ \\

\bottomrule
\end{tabular}
\endgroup
\end{table*}

\begingroup

We compare five draft tree control policies
under the same server verification scheduler.
The experiments use SpecBench under mixed networks
with $N=8$, covering G1 (Qwen3-14B / Qwen3-0.6B)
and G3 (Vicuna-7B / Vicuna-160M) in seeds 42-46.
All variants share the same base tree configuration:
$K=32$ candidate parents, depth $L=4$,
local branch width $W=16$, and node budget $B=32$.
Adaptive policies adjust their growth parameters
from this configuration.
Table~\ref{tab:ablation_study} reports the mean and
standard deviation between seeds for system performance
and mean client performance.
The Full CGSpec results are taken from the corresponding configurations in the main experiment.

\paragraph{Static Trees and Alternative Feedback Policies.}
\textit{Fixed tree, no JS} disables consistency feedback
and topology reshaping.
Limits expansion by the remaining node budget,
keeping tree size relatively stable.
\textit{Confidence prune} retains JS feedback
and computes confidence from the normalized entropy
of the draft model's Top-10 distribution.
When confidence falls below 0.6, it stops further
depth expansion.
The \textit{acceptance rate guide} uses verification acceptance
as feedback and adjusts the candidate branching threshold
through a controller, replacing the JS
control strategy.
These variants examine how fixed size control,
draft confidence guidance, and verification acceptance
feedback affect tree construction and end-to-end performance.

In G1, the fixed tree achieves a system throughput of
43.58 tokens/s.
Confidence pruning and acceptance rate guidance achieve
47.60 and 53.05 tokens/s, respectively, while full CGSpec
reaches 60.65 tokens/s.
The corresponding mean client throughput is
6.11, 7.03, 7.29, and 8.43 tokens/s.
In G3, confidence pruning achieves 53.52 tokens/s
of system throughput, below the fixed tree's
55.00 tokens/s.
Acceptance-rate guidance and full CGSpec achieve
61.14 and 65.71 tokens/s, respectively.
Full CGSpec achieves the best client throughput for both model pairs. For system throughput, it outperforms all alternatives in G1, while the JS budget only achieves a slightly higher value in G3.
\paragraph{Global Budget Scaling and Adaptive Tree Shape.}
\textit{JS budget only} and full CGSpec use the same
lightweight prefill configuration.
Their main difference lies in decoding:
The JS budget only proportionally scales the candidate-parent
count, local branch width, and node budget according
to the consistency state while using the default
beam search procedure.
Full CGSpec further coordinates budget, depth, and width,
favoring deeper and narrower exploration when consistency is high and shallower and
wider exploration when disagreement increases.
This comparison examines whether adaptive depth and width
allocation improves end-to-end performance beyond
global tree size adjustment.

In G1, full CGSpec increases system throughput from
52.08 to 60.65 tokens/s and mean client throughput
from 7.32 to 8.43 tokens/s relative to JS budget only.
For this model pair, consistency-driven tree size
scaling alone falls short of the full method's performance.
Further coordinating depth and width improves both
aggregate throughput and client progress.

G3 exhibits a different tradeoff.
The JS budget only achieves a mean system throughput of
66.53 tokens/s, compared with 65.71 tokens/s for full CGSpec.
Their mean client throughput is 9.98 and
10.95 tokens/s, respectively.
The full policy therefore achieves higher client
throughput, while global scaling alone has a slightly
higher mean system throughput.

\endgroup

\begin{figure*}[t]
    \centering
    \includegraphics[width=0.8\textwidth]
    {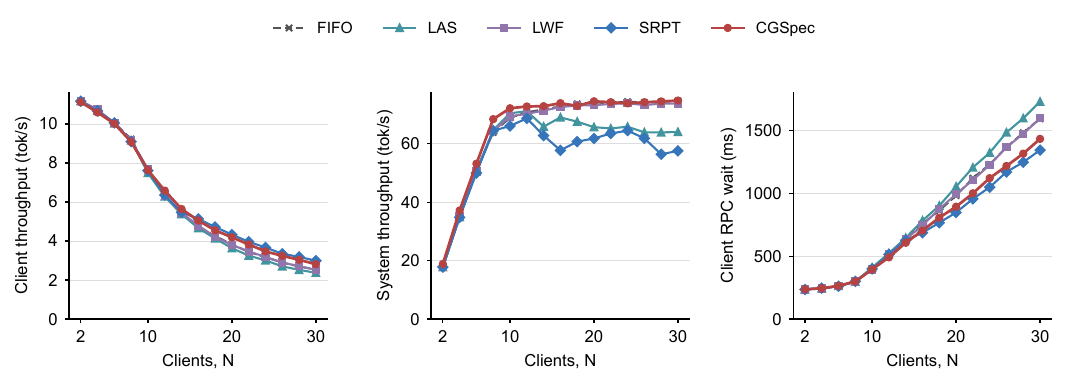}
    \caption{
    Comparison of five scheduling policies on
    G1/SpecBench under mixed networks.
    Left: mean client throughput.
    Middle: system throughput.
    Right: mean client RPC wait.
    }
    \label{fig:scheduling_analysis}
\end{figure*}

\subsection{Verification Scheduling Analysis}
\label{sec:scheduling-analysis}

\begingroup

We compare CGSpec with FIFO, LAS, LWF, and SRPT
to examine how different priority rules affect
client progress, aggregate throughput, and waiting time.
FIFO provides an arrival order reference.
LAS prioritizes clients with less accumulated token
progress, representing service based fairness.
LWF prioritizes clients with longer server side
waiting time, representing waiting time protection.
SRPT prioritizes requests according to estimated
serving cost in our implementation, representing
cost-aware prioritization.
Together, these policies cover arrival order,
attained service, waiting time, and estimated cost.
CGSpec combines estimated remaining work and acceptance
information with deadline and waiting time protection.

Figure~\ref{fig:scheduling_analysis} reports results
on G1 and SpecBench under mixed networks, with same requests in main test for $N=2,4,\ldots,30$ clients.

The three panels report mean client throughput,
system throughput, and client RPC wait.

As concurrency increases, client throughput decreases
and RPC waiting increases for all five policies,
while aggregate throughput exhibits different
saturation behavior.
At $N=30$, CGSpec achieves 2.81 tokens/s of client
throughput, compared with 2.53 for FIFO,
2.37 for LAS, and 2.52 for LWF.
Its system throughput reaches 74.75 tokens/s,
compared with 74.61, 64.00, and 73.72 tokens/s,
respectively.
Compared to the plotted FIFO reference, CGSpec
provides 11.1\% higher client throughput while
maintaining similar system throughput.
The mean client RPC wait decreases from 1596.1
to 1435.6 \,ms, a 10.1\% reduction.

SRPT exhibits a different tradeoff.
At $N=30$, it achieves the highest client throughput,
2.99 tokens/s, and the lowest RPC wait,
1345.9 \,ms, while its system throughput reaches
57.54 tokens /s.
CGSpec achieves 74.75 tokens /s of system throughput
at the same concurrency.
This comparison highlights the need to evaluate
client progress in conjunction with aggregate
verification throughput.
LWF remains close to FIFO in all three metrics,
while LAS exhibits lower throughput and longer
waiting at high concurrency.

At high concurrency, CGSpec combines high aggregate
throughput with improved client service
relative to FIFO.
Its scheduling rule accounts for both the expected
benefit of the next verification and the urgency
of serving a waiting client.
The resulting operating point balances aggregate
progress, client throughput, and waiting time as
competition for the shared verifier increases.

\endgroup

\section{Discussion}
\label{sec:discussion}

\begingroup

\textbf{Joint optimization of drafting and verification.}
The results indicate that efficient edge–cloud speculative decoding depends on the cost of the entire decoding cycle rather than drafting efficiency alone. Although CGSpec incurs additional local drafting time compared with current methods, it reduces communication, queueing, and verification time through consistency-guided tree construction and overlap-aware scheduling. This trade-off highlights the importance of evaluating speculative decoding based on end-to-end generation performance rather than isolated execution stages.

\textbf{Scheduling objectives under shared verification.}
The scheduling results reveal a trade-off between aggregate throughput and individual client progress. At high concurrency, SRPT achieves higher mean client throughput and lower RPC waiting time, whereas CGSpec maintains higher system throughput. FIFO also sustains competitive aggregate throughput but provides lower client throughput and longer waiting times than CGSpec. These results suggest that scheduling policies should account for both system efficiency and client service quality. By considering remaining verification work, acceptance information, and client return times, CGSpec provides a different operating point that favors aggregate progress while maintaining reasonable client-level performance.

\textbf{Deployment considerations.}
The relative importance of communication and queueing
depends on the model pair and network conditions.
Communication dominates several configurations under
weak networks, whereas server-side waiting remains
a major component for the larger target model.
Deployment should therefore consider client drafting
capacity, link characteristics, and verifier contention
together.
The current evaluation uses a single verification
request per service slot.
Extending the scheduler to batched verification
requires accounting for batch-dependent service time,
memory occupancy, and the interaction between
batch formation and client return times.
Multiple verifiers introduce an additional placement
decision alongside verification ordering.

There are also limitations in the current experimental evaluation. First, network conditions are emulated by injecting controlled delays based on representative 4G- and 5G-like profiles, rather than measured through real cellular connections. This setup enables repeatable comparisons under identical network conditions and helps isolate the effects of different scheduling policies. However, it does not fully capture time-varying wireless bandwidth, packet loss, handover events, and other dynamics of operational mobile networks. Second, our experiments use NVIDIA A40 GPUs as clients and AMD Instinct MI250 GPUs as cloud verification servers. Although this setup allows us to examine heterogeneous client–server execution and shared-verifier contention, it does not characterize the additional constraints of mobile and resource-constrained edge devices, such as limited memory capacity, power consumption, and thermal constraints. Evaluating CGSpec on physical mobile networks and resource-constrained edge platforms is an important direction for future work. In addition, the current evaluation uses a homogeneous GPU configuration across clients, and the impact of heterogeneous client-side computing capabilities remains to be investigated.
\endgroup

\section{Conclusion}
\label{sec:conclusion}

\begingroup

We presented CGSpec, a consistency-guided framework for multi-client edge–cloud speculative decoding. CGSpec uses draft–target distribution divergence to regulate speculative tree growth and a nomination and competition strategy to select candidate tokens under a limited budget. Its overlap-aware scheduler accounts for remaining verification work and client return times when allocating shared verification resources.

Experiments across four model pairs, three datasets, and heterogeneous network conditions show that CGSpec achieves the highest mean system and client throughput among the evaluated methods at both 8 and 16 clients. Compared with current SOTA methods, it improves system throughput by 12.2\% and 10.2\%, and mean client throughput by 14.0\% and 33.0\%, respectively. Further analyses examine its speculative efficiency, scheduling behavior, scalability, and latency distribution.

These results demonstrate the benefits of coordinating speculative tree construction with verification scheduling in multi client edge–cloud inference. They also provide insights into managing shared verification resources under heterogeneous network conditions.
\endgroup


\ifCLASSOPTIONcaptionsoff
  \newpage
\fi



%

\bibliographystyle{IEEEtran}
\bibliography{software}

@inproceedings{leviathan2023specdec,
  title = {Fast Inference from Transformers via Speculative Decoding},
  author = {Leviathan, Yaniv and Kalman, Matan and Matias, Yossi},
  booktitle = {Proceedings of the 40th International Conference on Machine Learning},
  pages = {19274--19286},
  year = {2023},
  volume = {202},
  publisher = {PMLR},
  }

@misc{chen2023specsampling,
  title = {Accelerating Large Language Model Decoding with Speculative Sampling},
  author = {Chen, Charlie and Borgeaud, Sebastian and Irving, Geoffrey and Lespiau, Jean-Baptiste and Sifre, Laurent and Jumper, John},
  year = {2023},
  eprint = {2302.01318},
  archivePrefix = {arXiv},
  primaryClass = {cs.CL},
  url = {https://arxiv.org/abs/2302.01318}
}

@inproceedings{cai2024medusa,
  title = {Medusa: Simple {LLM} Inference Acceleration Framework with Multiple Decoding Heads},
  author = {Cai, Tianle and Li, Yuhong and Geng, Zhengyang and Peng, Hongwu and Lee, Jason D. and Chen, Deming and Dao, Tri},
  booktitle = {Proceedings of the 41st International Conference on Machine Learning},
  pages = {5209--5235},
  year = {2024},
  volume = {235},
  publisher = {PMLR},
}

@article{li2026eagle,
  title={Eagle-3: Scaling up inference acceleration of large language models via training-time test},
  author={Li, Yuhui and Wei, Fangyun and Zhang, Chao and Zhang, Hongyang},
  journal={Advances in Neural Information Processing Systems},
  volume={38},
  pages={136737--136756},
  year={2026}
}

@article{hu2026adaspec,
  title={Adaspec: Selective knowledge distillation for efficient speculative decoders},
  author={Hu, Yuezhou and Guo, Jiaxin and Feng, Xinyu and Zhao, Tuo},
  journal={Advances in Neural Information Processing Systems},
  volume={38},
  pages={88736--88758},
  year={2026}
}

@misc{miao2024specinfer,
  title = {SpecInfer: Accelerating Generative Large Language Model Serving with Tree-based Speculative Inference and Verification},
  author = {Miao, Xupeng and Oliaro, Gabriele and Zhang, Zhihao and Cheng, Xinhao and Wang, Zeyu and Zhang, Zhengxin and Wong, Rae Ying Yee and Zhu, Alan and Yang, Lijie and Shi, Xiaoxiang and Shi, Chunan and Chen, Zhuoming and Arfeen, Daiyaan and Abhyankar, Reyna and Jia, Zhihao},
  year = {2024},
  eprint = {2305.09781},
  archivePrefix = {arXiv},
  primaryClass = {cs.CL},
  doi = {10.1145/3620666.3651335},
  url = {https://arxiv.org/abs/2305.09781}
}

@misc{spector2023staged,
  title = {Accelerating {LLM} Inference with Staged Speculative Decoding},
  author = {Spector, Benjamin and Re, Chris},
  year = {2023},
  eprint = {2308.04623},
  archivePrefix = {arXiv},
  primaryClass = {cs.AI},
  url = {https://arxiv.org/abs/2308.04623}
}

@misc{chen2025sequoia,
  title = {Sequoia: Scalable, Robust, and Hardware-aware Speculative Decoding},
  author = {Chen, Zhuoming and May, Avner and Svirschevski, Ruslan and Huang, Yuhsun and Ryabinin, Max and Jia, Zhihao and Chen, Beidi},
  year = {2025},
  eprint = {2402.12374},
  archivePrefix = {arXiv},
  primaryClass = {cs.CL},
  url = {https://arxiv.org/abs/2402.12374}
}

@misc{li2024eagle2,
  title = {{EAGLE-2}: Faster Inference of Language Models with Dynamic Draft Trees},
  author = {Li, Yuhui and Wei, Fangyun and Zhang, Chao and Zhang, Hongyang},
  year = {2024},
  eprint = {2406.16858},
  archivePrefix = {arXiv},
  primaryClass = {cs.CL},
  url = {https://arxiv.org/abs/2406.16858}
}

@misc{mamou2024disco,
  title = {Dynamic Speculation Lookahead Accelerates Speculative Decoding of Large Language Models},
  author = {Mamou, Jonathan and Pereg, Oren and Korat, Daniel and Berchansky, Moshe and Timor, Nadav and Wasserblat, Moshe and Schwartz, Roy},
  year = {2024},
  eprint = {2405.04304},
  archivePrefix = {arXiv},
  primaryClass = {cs.CL},
  url = {https://arxiv.org/abs/2405.04304}
}

@inproceedings{li2025sled,
  title={Sled: A speculative llm decoding framework for efficient edge serving},
  author={Li, Xiangchen and Spatharakis, Dimitrios and Ghafouri, Saeid and Fan, Jiakun and Vandierendonck, Hans and John, Deepu and Ji, Bo and Nikolopoulos, Dimitrios S},
  booktitle={Proceedings of the Tenth ACM/IEEE Symposium on Edge Computing},
  pages={1--8},
  year={2025}
}

@article{park2026specedge,
  title={SpecEdge: Scalable edge-assisted serving framework for interactive LLMs},
  author={Park, Jinwoo and Cho, Seunggeun and Han, Dongsu},
  journal={Advances in Neural Information Processing Systems},
  volume={38},
  pages={92668--92694},
  year={2026}
}

@inproceedings{tran2025goodspeed,
  title={GoodSpeed: Optimizing Fair Goodput with Adaptive Speculative Decoding in Distributed Edge Inference},
  author={Tran, Phuong and Liu, Tzu-Hao and Le, Long Tan and Nguyen, Tung-Anh and La, Van Quan and Yu, Eason and Shu, Han and Hong, Choong Seon and Tran, Nguyen H},
  booktitle={IEEE INFOCOM 2026-IEEE Conference on Computer Communications},
  pages={1--10},
  year={2026},
  organization={IEEE}
}

@misc{yu2025dsd,
  title = {{DSD}: A Distributed Speculative Decoding Solution for Edge-Cloud Agile Large Model Serving},
  author = {Yu, Fengze and Li, Leshu and McDanel, Brad and Zhang, Sai Qian},
  year = {2025},
  eprint = {2511.21669},
  archivePrefix = {arXiv},
  primaryClass = {cs.LG},
  url = {https://arxiv.org/abs/2511.21669}
}

@article{zhang2026qsv,
  title={Quantize-sample-and-verify: LLM acceleration via adaptive edge-cloud speculative decoding},
  author={Zhang, Guangyi and Cai, Yunlong and Yu, Guanding and Popovski, Petar and Simeone, Osvaldo},
  journal={IEEE Communications Letters},
  volume={30},
  pages={852--856},
  year={2026},
  publisher={IEEE}
}

@misc{bhattacharjee2025conformal,
  title = {Conformal Sparsification for Bandwidth-Efficient Edge-Cloud Speculative Decoding},
  author = {Bhattacharjee, Payel and Tian, Fengwei and Zhong, Meiyu and Zhang, Guangyi and Simeone, Osvaldo and Tandon, Ravi},
  year = {2025},
  eprint = {2510.09942},
  archivePrefix = {arXiv},
  primaryClass = {cs.LG},
  url = {https://arxiv.org/abs/2510.09942}
}

@inproceedings{xia2024survey,
  title = {Unlocking Efficiency in Large Language Model Inference: A Comprehensive Survey of Speculative Decoding},
  author = {Xia, Heming and Yang, Zhe and Dong, Qingxiu and Wang, Peiyi and Li, Yongqi and Ge, Tao and Liu, Tianyu and Li, Wenjie and Sui, Zhifang},
  booktitle = {Findings of the Association for Computational Linguistics: {ACL} 2024},
  pages = {7655--7671},
  year = {2024},
  doi = {10.18653/v1/2024.findings-acl.456},
}

@misc{li2024eagle,
  title = {{EAGLE}: Speculative Sampling Requires Rethinking Feature Uncertainty},
  author = {Li, Yuhui and Wei, Fangyun and Zhang, Chao and Zhang, Hongyang},
  year = {2024},
  eprint = {2401.15077},
  archivePrefix = {arXiv},
  primaryClass = {cs.CL},
  url = {https://arxiv.org/abs/2401.15077}
}

@article{xiong2025dyspec,
  title = {DySpec: Faster Speculative Decoding with Dynamic Token Tree Structure},
  author = {Xiong, Yunfan and Zhang, Ruoyu and Li, Yanzeng and Zou, Lei},
  journal = {World Wide Web},
  volume = {28},
  number = {3},
  year = {2025},
  month = may,
  doi = {10.1007/s11280-025-01344-0},
}

@article{guan2026yggdrasil,
  title={Yggdrasil: Bridging dynamic speculation and static runtime for latency-optimal tree-based llm decoding},
  author={Guan, Yue and Yu, Changming and Fang, Shihan and Hu, Weiming and Pan, Zaifeng and Wang, Zheng and Liu, Zihan and Zhou, Yangjie and Ding, Yufei and Guo, Minyi and others},
  journal={Advances in Neural Information Processing Systems},
  volume={38},
  pages={86706--86726},
  year={2026}
}

@misc{ning2025dssd,
  title = {{DSSD}: Efficient Edge-Device {LLM} Deployment and Collaborative Inference via Distributed Split Speculative Decoding},
  author = {Ning, Jiahong and Zheng, Ce and Yang, Tingting},
  year = {2025},
  eprint = {2507.12000},
  archivePrefix = {arXiv},
  primaryClass = {eess.SP},
  url = {https://arxiv.org/abs/2507.12000}
}

@inproceedings{ye2025jupiter,
  title={Jupiter: Fast and resource-efficient collaborative inference of generative llms on edge devices},
  author={Ye, Shengyuan and Ouyang, Bei and Zeng, Liekang and Qian, Tianyi and Chu, Xiaowen and Tang, Jian and Chen, Xu},
  booktitle={IEEE INFOCOM 2025-IEEE Conference on Computer Communications},
  pages={1--10},
  year={2025},
  organization={IEEE}
}

@misc{liu2025flowspec,
  title = {FlowSpec: Continuous Pipelined Speculative Decoding for Efficient Distributed {LLM} Inference},
  author = {Liu, Xing and Luo, Lizhuo and Tang, Ming and Huang, Chao and Chen, Xu},
  year = {2025},
  eprint = {2507.02620},
  archivePrefix = {arXiv},
  primaryClass = {cs.DC},
  url = {https://arxiv.org/abs/2507.02620}
}

@inproceedings{mcdanel2024amusd,
  title={Amusd: Asynchronous multi-device speculative decoding for llm acceleration},
  author={McDanel, Bradley},
  booktitle={2025 IEEE International Symposium on Circuits and Systems (ISCAS)},
  pages={1--5},
  year={2025},
  organization={IEEE}
}

@article{li2026wisp,
  title={WISP: waste-and interference-suppressed distributed speculative LLM serving at the edge via dynamic drafting and SLO-aware batching},
  author={Li, Xiangchen and Fan, Jiakun and Wang, Qingyuan and Spatharakis, Dimitrios and Ghafouri, Saeid and Vandierendonck, Hans and John, Deepu and Ji, Bo and Butt, Ali R and Nikolopoulos, Dimitrios},
  journal={Proceedings of the ACM on Measurement and Analysis of Computing Systems},
  volume={10},
  number={2},
  pages={1--31},
  year={2026},
  publisher={ACM, New York, NY, USA}
}

@ARTICLE{10812936,
  author={Xu, Daliang and Yin, Wangsong and Zhang, Hao and Jin, Xin and Zhang, Ying and Wei, Shiyun and Xu, Mengwei and Liu, Xuanzhe},
  journal={IEEE Transactions on Mobile Computing}, 
  title={EdgeLLM: Fast On-Device LLM Inference With Speculative Decoding}, 
  year={2025},
  volume={24},
  number={4},
  pages={3256-3273},
  doi={10.1109/TMC.2024.3513457}}

@inproceedings{ghoshal2023performance,
  title={Performance of cellular networks on the wheels},
  author={Ghoshal, Moinak and Khan, Imran and Kong, Z Jonny and Dinh, Phuc and Meng, Jiayi and Hu, Y Charlie and Koutsonikolas, Dimitrios},
  booktitle={Proceedings of the 2023 ACM on Internet Measurement Conference},
  pages={678--695},
  year={2023}
}

@article{lim2023empirical,
  title={An empirical study of 5G: Effect of edge on transport protocol and application performance},
  author={Lim, Hyoyoung and Lee, Jinsung and Lee, Jongyun and Sathyanarayana, Sandesh Dhawaskar and Kim, Junseon and Nguyen, Anh and Kim, Kwang Taik and Im, Youngbin and Chiang, Mung and Grunwald, Dirk and others},
  journal={IEEE Transactions on Mobile Computing},
  volume={23},
  number={4},
  pages={3172--3186},
  year={2023},
  publisher={IEEE}
}

@inproceedings{xu2021cloud,
  title={From cloud to edge: a first look at public edge platforms},
  author={Xu, Mengwei and Fu, Zhe and Ma, Xiao and Zhang, Li and Li, Yanan and Qian, Feng and Wang, Shangguang and Li, Ke and Yang, Jingyu and Liu, Xuanzhe},
  booktitle={Proceedings of the 21st ACM internet measurement conference},
  pages={37--53},
  year={2021}
}

@article{yang2025qwen3,
  title={Qwen3 technical report},
  author={Yang, An and Li, Anfeng and Yang, Baosong and Zhang, Beichen and Hui, Binyuan and Zheng, Bo and Yu, Bowen and Gao, Chang and Huang, Chengen and Lv, Chenxu and others},
  journal={arXiv preprint arXiv:2505.09388},
  year={2025}
}

@article{JMLR:v21:20-074,
  author  = {Colin Raffel and Noam Shazeer and Adam Roberts and Katherine Lee and Sharan Narang and Michael Matena and Yanqi Zhou and Wei Li and Peter J. Liu},
  title   = {Exploring the Limits of Transfer Learning with a Unified Text-to-Text Transformer},
  journal = {Journal of Machine Learning Research},
  year    = {2020},
  volume  = {21},
  number  = {140},
  pages   = {1--67},
}

@article{merity2016pointer,
  title={Pointer sentinel mixture models},
  author={Merity, Stephen and Xiong, Caiming and Bradbury, James and Socher, Richard},
  journal={arXiv preprint arXiv:1609.07843},
  year={2016}
}

@article{ubicomp513,
author = {Huang, Yifei and Xu, Jilan and Pei, Baoqi and Yang, Lijin and Zhang, Mingfang and He, Yuping and Chen, Guo and Chen, Xinyuan and Wang, Yaohui and Nie, Zheng and Liu, Jinyao and Lin, Dechen and Fang, Fang and Li, Kunpeng and Yuan, Chang and Qiao, Yu and Wang, Yali and Wang, Limin},
title = {Vinci: A Real-time Smart Assistant Based on Egocentric Vision-language Model for Portable Devices},
year = {2025},
issue_date = {September 2025},
publisher = {Association for Computing Machinery},
address = {New York, NY, USA},
volume = {9},
number = {3},
doi = {10.1145/3749513},
journal = {Proc. ACM Interact. Mob. Wearable Ubiquitous Technol.},
month = sep,
articleno = {88},
numpages = {33}
}

@article{edmonds2005maiden,
  title={A maiden analysis of longest wait first},
  author={Edmonds, Jeff and Pruhs, Kirk},
  journal={ACM Transactions on Algorithms (TALG)},
  volume={1},
  number={1},
  pages={14--32},
  year={2005},
  publisher={ACM, New York, NY, USA}
}

@inproceedings{rai2003analysis,
  title={Analysis of LAS scheduling for job size distributions with high variance},
  author={Rai, Idris A and Urvoy-Keller, Guillaume and Biersack, Ernst W},
  booktitle={Proceedings of the 2003 ACM SIGMETRICS international conference on Measurement and modeling of computer systems},
  pages={218--228},
  year={2003}
}

@article{smith1978new,
  title={A new proof of the optimality of the shortest remaining processing time discipline},
  author={Smith, Donald R},
  journal={Operations Research},
  volume={26},
  number={1},
  pages={197--199},
  year={1978},
  publisher={INFORMS}
}


\end{document}